# Expectation-induced modulation of metastable activity underlies faster coding of sensory stimuli

L. Mazzucato[1,2], G. La Camera[1,3],*, A. Fontanini[1,3],*

[1] *Department of Neurobiology and Behavior, State University of New York at Stony Brook, Stony Brook, NY 11794*

[2] *Departments of Biology and Mathematics and Institute of Neuroscience, University of Oregon, Eugene, OR 97403*

[3] *Graduate Program in Neuroscience, State University of New York at Stony Brook, Stony Brook, NY 11794*

* Co-senior authors

Running Title: Metastable activity mediates expectation

**Contact:**
Alfredo Fontanini[a] and Giancarlo La Camera[b]

[a] Department of Neurobiology & Behavior, Life Sciences Building 516
State University of New York at Stony Brook, Stony Brook, NY 11794
alfredo.fontanini@stonybrook.edu

[b] Department of Neurobiology & Behavior, Life Sciences Building 513
State University of New York at Stony Brook, Stony Brook, NY 11794
giancarlo.lacamera@stonybrook.edu

## Abstract

Sensory stimuli can be recognized more rapidly when they are expected. This phenomenon depends on expectation affecting the cortical processing of sensory information. However, virtually nothing is known on the mechanisms responsible for the effects of expectation on sensory networks. Here, we report a novel computational mechanism underlying the expectation-dependent acceleration of coding observed in the gustatory cortex (GC) of alert rats. We use a recurrent spiking network model with a clustered architecture capturing essential features of cortical activity, including the metastable activity observed in GC before and after gustatory stimulation. Relying both on network theory and computer simulations, we propose that expectation exerts its function by modulating the intrinsically generated dynamics preceding taste delivery. Our model, whose predictions are confirmed in the experimental data, demonstrates how the modulation of intrinsic metastable activity can shape sensory coding and mediate cognitive processes such as the expectation of relevant events. Altogether, these results provide a biologically plausible theory of expectation and ascribe a new functional role to intrinsically generated, metastable activity.

## Introduction

Expectation exerts a strong influence on sensory processing. It improves stimulus detection, enhances discrimination between multiple stimuli and biases perception towards an anticipated stimulus[1-3]. These effects, demonstrated experimentally for various sensory modalities and in different species[2,4-7], can be attributed to changes in sensory processing occurring in primary sensory cortices. However, despite decades of investigations, little is known regarding how expectation shapes the cortical processing of sensory information.

While different forms of expectation likely rely on a variety of neural mechanisms, modulation of pre-stimulus activity is believed to be a common underlying feature[8-10]. Here, we investigate the link between pre-stimulus activity and the phenomenon of general expectation in a recent set of experiments performed in gustatory cortex (GC) of alert rats[6]. In those experiments, rats were trained to expect the intraoral delivery of one of four possible tastants following an anticipatory cue. The use of a single cue allowed the animal to predict the availability of gustatory stimuli, without forming expectations on which specific taste was being delivered. Cues predicting the general availability of taste modulated the firing rates of GC neurons. Tastants delivered after the cue were encoded more rapidly than uncued tastants, and this improvement was phenomenologically attributed to the activity evoked by the preparatory cue. However, the precise computational mechanism linking faster coding of taste and cue responses remains unknown.

Here we propose a mechanism whereby an anticipatory cue modulates the timescale of temporal dynamics in a recurrent population model of spiking neurons. In the model proposed here, neurons are organized in strongly connected clusters and produce sequences of metastable states similar to those observed during both pre-stimulus and evoked activity periods[11-17]. A metastable state is a vector of firing rates across simultaneously recorded neurons that can last for several hundred milliseconds before giving way to the next state in a sequence. The ubiquitous presence of state sequences in many cortical areas and behavioral contexts[18-24] has raised the issue of their role in sensory and cognitive processing. Here, we elucidate the central role played by pre-stimulus metastable states in processing forthcoming stimuli, and show how cue-induced modulations of state sequences drive anticipatory coding. Specifically, we show that an anticipatory cue affects sensory coding by decreasing the duration of metastable states and accelerating the pace of state sequences. This phenomenon, which results from a reduction in the effective energy barriers separating the metastable states, accelerates the onset of specific states coding for the presented stimulus, thus mediating the effects of general expectation. The predictions of our model were confirmed in a new analysis of the experimental data, also reported here.

Altogether, our results provide a model for general expectation, based on the modulation of pre-stimulus ongoing cortical dynamics by anticipatory cues, leading to acceleration of sensory coding.

## Results

**Anticipatory cue accelerates stimulus coding in a clustered population of neurons**

To uncover the computational mechanism linking cue-evoked activity with coding speed, we modeled the gustatory cortex (GC) as a population of recurrently connected excitatory and inhibitory spiking neurons. In this model, excitatory neurons are arranged in clusters[12,25,26] (Fig. 1a), reflecting the existence of assemblies of functionally correlated neurons in GC and other cortical areas[27,28]. Recurrent synaptic weights between neurons in the same cluster are potentiated compared to neurons in different clusters, to account for metastability in GC[11,16] and in keeping with evidence from electrophysiological and imaging experiments[27,28] [29,30]. This spiking network also has bidirectional random and homogeneous (i.e., non-clustered) connections among inhibitory neurons and between inhibitory and excitatory neurons. Such connections stabilize network activity by preventing runaway excitation and play a role in inducing the observed metastability[11,12,15].

The model was probed by sensory inputs modeled as depolarizing currents injected into randomly selected neurons. We used four sets of simulated stimuli, wired to produce gustatory responses reminiscent of those observed in the experiments in the presence of sucrose, sodium chloride, citric acid, and quinine (see Supplementary Methods for details). The specific connectivity pattern used was inferred by the presence of both broadly and narrowly tuned responses in GC[31,32], and the temporal dynamics of the inputs were varied to determine the robustness of the model (Supplementary Results, Sec. 1.3).

In addition to input gustatory stimuli, we included anticipatory inputs designed to produce cue-responses analogous to those seen experimentally in the case of general expectation. To simulate general expectation, we connected anticipatory inputs with random neuronal targets in the network. The peak value of the cue-induced current for each neuron was sampled from a normal distribution with zero mean and fixed variance (see Fig. S1-S2 and Supplementary Results for details), thus introducing a spatial variance in the afferent currents. This choice reflected the large heterogeneity of cue responses observed in the empirical data, where excited and inhibited neural responses occurred in similar proportions[10] and overlapped partially with taste responses[6,10]. Fig 1b shows two representative cue-responsive neurons in the model: one inhibited by the cue and one excited by the cue (more details and examples are reported in the Supplementary Results).

Given these conditions, we simulated the experimental paradigm adopted in awake-behaving rats to demonstrate the effects of general expectation[6,10]. In the original experiment, rats were trained to self-administer into an intra-oral cannula one of four possible tastants following an anticipatory cue. At random trials and time during the inter-trial interval, tastants were unexpectedly delivered in the absence of a cue. To match this experiment, the simulated paradigm interleaves two conditions: in expected trials, a stimulus (out of 4) is delivered at *t=0* after an anticipatory cue (the same for all stimuli) delivered at *t=-0.5*s (Fig. 1b); in unexpected trials the same stimuli are presented in the absence of the cue. Importantly, in the general expectation paradigm adopted here, the anticipatory cue is identical for all stimuli in the expected condition. Therefore, it does not convey any information regarding the identity of the stimulus being delivered.

We tested whether cue presentation affected stimulus coding. A multi-class classifier (see Methods and Fig. S3) was used to assess the information about the stimuli encoded in the neural activity, where the four class labels correspond to the four tastants. Stimulus identity was encoded well in both conditions, reaching perfect average accuracy across the four tastants after a few hundred milliseconds (Fig. 1c, across-taste average decoding accuracy). However, comparing the time course of the decoding accuracy between conditions, we found that the increase in decoding accuracy was significantly faster in expected than in unexpected trials (Fig. 1c, pink and blue curves represent expected and unexpected conditions, respectively). Indeed, the onset time of a significant decoding occurred earlier in the expected vs. the unexpected condition (decoding latency was $0.13 \pm 0.01$ s [mean$\pm$s.e.m.] for expected compared to $0.21 \pm 0.02$ s for unexpected, across 20 independent sessions; p=0.002, signed-rank=14, d.o.f.=39; inset in Fig. 1c). Similar decoding accuracies were obtained for each individual tastant separately (see Supplementary Results and Fig. S3). Thus, in the model network, the interaction of cue response and activity evoked by the stimuli results in faster encoding of the stimuli themselves, mediating the expectation effect.

To clarify the role of neural clusters in mediating expectation, we simulated the same experiments in a homogeneous network (i.e., without clusters) operating in the balanced asynchronous regime[25,26] (Fig. 1d, intra- and inter-cluster weights were set equal, all other network parameters and inputs were the same as for the clustered network). Even though single neurons' responses to the anticipatory cue were comparable to the ones observed in the clustered network (Fig. 1e, Fig. S2 and Supplementary Results), stimulus encoding was not affected by cue presentation (Fig. 1f). In particular, the onset of a significant decoding was similar in the two conditions (latency of significant decoding was $0.17 \pm 0.01$ s for expected and $0.16 \pm 0.01$ s for unexpected tastes averaged across 20 sessions; p=0.31, signed-rank=131, d.o.f.=39; inset in Fig. 1f).

The anticipatory activity observed in the clustered network was robust to variations in key parameters related to the sensory and anticipatory inputs, as well as network connectivity, size and architecture (Fig. S4-6 and Supplementary Results). Furthermore, acceleration of coding depended on the patterns of connectivity of anticipatory inputs, specifically on the fact that it increased the spatial variance in the cue afferent currents (Fig. S6). In a model where the cue recruited the recurrent inhibition (by increasing the input currents to the inhibitory population), stimulus coding was decelerated (Fig. S7), suggesting a potential mechanism mediating the effect of distractors.

Overall, these results demonstrate that a clustered network of spiking neurons can successfully reproduce the acceleration of sensory coding induced by expectation and that removing clustering impairs this function.

**Anticipatory cue speeds up the network's dynamics**

Having established that a clustered architecture mediates the effects of expectation on coding, we investigated the underlying mechanism.

Clustered networks spontaneously generate highly structured activity characterized by coordinated patterns of ensemble firing. This activity results from the network hopping between metastable states in which different combinations of clusters are simultaneously activated[11,14,15]. To understand how anticipatory inputs affected network dynamics, we analyzed the effects of cue presentation for a prolonged period of 5 seconds in the absence of stimuli. Activating anticipatory inputs led to changes in network dynamics, with clusters turning on and off more frequently in the presence of the cue (Fig. 2a). We quantified this effect by showing that a cue-induced increase in input spatial variance ($\sigma^2$) led to a shortened cluster activation lifetime (top panel in Fig. 2b; Kruskal-Wallis one-way ANOVA: $p<10^{-17}$, $\chi^2(5)=91.2$), and a shorter cluster inter-activation interval (i.e., quiescent intervals between consecutive activations of the same cluster, bottom panel in Fig. 2b, kruskal-wallis one-way ANOVA: $p<10^{-18}$, $\chi^2(5)=98.6$).

Previous work has demonstrated that metastable states of co-activated clusters result from attractor dynamics[11,14,15]. Hence, the shortening of cluster activations and inter-activation intervals observed in the model could be due to modifications in the network's attractor dynamics. To test this hypothesis, we performed a mean field theory analysis[33-36] of a simplified network with only two clusters, therefore producing a reduced repertoire of configurations. Those include two configurations in which either cluster is active and the other inactive ('A' and 'B' in Fig. 2c), and a configuration where both clusters are moderately active ('C'). The dynamics of this network can be analyzed using a reduced, self-consistent theory of a single excitatory cluster, said to be *in focus*[33] (see Methods for details), based on the effective transfer function relating the input and output firing rates of the cluster ($r$ and $r_{out}$, Fig. 2c). The latter are equal in the A, B and C network configurations described above – also called 'fixed points' since these are the points where the transfer function intersects the identity line, $r_{out} = \Phi(r_{in})$.

Configurations A and B would be stable in an infinitely large network, but they are only metastable in networks of finite size, due to intrinsically generated variability[15]. Transitions between metastable states can be modeled as a diffusion process and analyzed with Kramers' theory[37], according to which the transition rates depend on the height Δ of an effective energy barrier separating them[15,37]. In our theory, the effective energy barriers (Fig. 2c, bottom row) are obtained as the area of the region between the identity line and the transfer function (shaded areas in top row of Fig. 2c; see Methods for details). The effective energy is constructed so that its local minima correspond to stable fixed points (here, A and B) while local maxima correspond to unstable fixed points (C). Larger barriers correspond to less frequent transitions between stable configurations, whereas lower barriers increase the transition rates and therefore accelerate the network's metastable dynamics.

This picture provides the substrate for understanding the role of the anticipatory cue in the expectation effect. Basically, the presentation of the cue modulates the shape of the effective transfer function, which results in the reduction of the effective energy barriers. More specifically, the cue-induced increase in the spatial variance, $\sigma^2$, of the afferent current flattens the transfer function along the identity line, reducing the area between the two (shaded regions in Fig. 2c). In turn, this reduces the effective energy barrier separating the two configurations (Fig. 2c, bottom row), resulting in faster dynamics. The larger the cue-induced spatial variance $\sigma^2$ in the afferent currents, the faster the dynamics (Fig. 2d; lighter shades represent larger $\sigma$s).

In summary, this analysis shows that the anticipatory cue increases the spontaneous transition rates between the network's metastable configurations by reducing the effective energy barrier necessary to hop among configurations. In the following we uncover an important consequence of this phenomenon for sensory processing.

**Anticipatory cue induces faster onset of taste-coding states**

The cue-induced modulation of attractor dynamics led us to formulate a hypothesis for the mechanism underlying the acceleration of coding: The activation of anticipatory inputs prior to sensory stimulation may allow the network to enter more easily configurations encoding stimuli while exiting more easily non-coding configurations. Fig. 3a shows simulated population rasters in response to the same stimulus presented in the absence of a cue or after a cue. Spikes in red hue represent activity in taste-selective clusters and show a faster activation latency in response to the stimulus preceded by the cue compared to the uncued condition. A systematic analysis revealed that in the cued condition, the clusters activated by the subsequent stimulus had a significantly faster activation latency than in the uncued condition (Fig. 3b, $0.22 \pm 0.01$ s (mean$\pm$s.e.m.) during cued compared to $0.32 \pm 0.01$ s for uncued stimuli; $p<10^{-5}$, rank sum test R(39)=232).

We elucidated this effect using mean field theory. In the simplified two-cluster network of Fig. 3c (the same network as in Fig. 2d), the configuration where the taste-selective cluster is active ("coding state") and the nonselective cluster is active ("non-coding state") have initially the same effective potential energy, in the absence of stimulation (local minima of the black line in Fig. 3c), separated by an effective energy barrier whose height is reduced by the anticipatory cue (dashed vs. full line). When the taste stimulus is presented, it activates the stimulus-selective cluster, so that the coding state will now sit in a deeper well (lighter lines) compared to non-coding state. Stronger stimuli (lighter shades in Fig. 3c) increase the difference between the wells' depths breaking their initial symmetry, so that now a transition from the non-coding to the coding state is more likely than a transition from the coding to the non-coding state[37], (also in the absence of the cue; full lines). The anticipatory cue reduces further the existing barrier and thereby increases the transition rate into coding configurations. This results into *faster* coding, on average, of the stimuli encoded by those states.

We tested this model prediction on the data from Samuelsen et al. (Fig 4)[6]. To compare the data to the model simulations, we randomly sampled ensembles of model neurons so as to match the sizes of the empirical datasets. Since we only have access to a subset of neurons in the experiments, rather than the full network configuration, we segmented the ensemble activity in sequences of metastable states via a Hidden Markov Model (HMM) analysis (see Methods). Previous work has demonstrated that HMM states, i.e. patterns of coordinated ensemble firing activity, can be treated as proxies of metastable network configurations[11]. In particular, activation of taste-coding configurations for a particular stimulus results in HMM states containing information about that stimulus (i.e., taste-coding HMM states). If the hypothesis originating from the model is correct, transitions from non-coding HMM states to taste-coding HMM states should be faster in the presence of the cue compared to uncued trials. We indeed found faster transitions to HMM coding states in cued trials for both model and data (Fig. 4a and 4c, respectively; color-coded horizontal bars overlay coding states). The latency of coding states was significantly faster during cued compared to uncued trials in both the model (Fig. 4b, mean latency of the first coding state was $0.32 \pm 0.02$ s for expected vs $0.38 \pm 0.01$ s for unexpected trials; rank sum test R(39)=319, p=0.014) and the empirical data (Fig. 4d: $0.46 \pm 0.02$ s for expected vs $0.56 \pm 0.03$ s for unexpected trials; rank sum test R(37)=385, p=0.026).

Altogether, these results demonstrate that anticipatory inputs speed up sensory coding by reducing the effective energy barriers from non-coding to coding metastable states (i.e., the transitions facilitated by the stimulus).

## Discussion

Expectations modulate perception and sensory processing. Typically, expected stimuli are recognized more accurately and rapidly than unexpected ones [1-3]. In the gustatory cortex, acceleration of taste coding has been related to changes in firing activity evoked by cues predicting the general availability of tastants[6]. However, the computational mechanisms linking pre-stimulus activity with changes in the latency of sensory coding are still unknown. Here we propose a novel mechanism that explains the effects of expectation through the modulation of the dynamics intrinsically generated by the cortex. Our results provide a new functional interpretation for the intrinsically generated activity that is ubiquitously observed in cortical circuits[11,18-21,24,38-42].

The proposed mechanism requires a recurrent spiking network where excitatory neurons are arranged in clusters, which has been demonstrated to capture essential features of the dynamics of neural activity in sensory circuits[11,16]. In such a model, network activity during both spontaneous and stimulus-evoked periods unfolds through state sequences, each state representing a metastable network attractor. In response to an anticipatory cue, the pace of state sequences speeds up, accelerated by a higher transition probability among states. The latter is caused by lowering the potential barrier separating metastable states in the attractor landscape. This anticipates the offset of states not conveying taste information and the onset of states containing the most information about the delivered stimulus ('coding states'), causing the faster decoding observed by Samuelsen *et al*[6] (see Fig. 1c).

Notably, this novel mechanism for anticipation is unrelated to increases in network excitability which would lead to unidirectional changes in taste-evoked firing rates. It relies instead on an increase in the spatial variance of the cue afferent currents to the sensory network brought about by the anticipatory cue. This increase in the input's variance is observed experimentally after training[10], and is therefore the consequence of having learned the anticipatory meaning of the cue. The acceleration of the dynamics of state sequences predicted by the model was also confirmed in the data from ensembles of simultaneously recorded neurons in awake-behaving rats.

These results provide a precise explanatory link between the intrinsic dynamics of neural activity in a sensory circuit and a specific cognitive process, that of general expectation[6] (see also [43,44]).

**Clustered connectivity and metastable states**

A key feature of our model is the clustered architecture of the excitatory population. Removing excitatory clusters eliminates the cue-induced anticipatory effect (Fig. 1d-f). Theoretical work in recurrent networks had previously shown that a clustered architecture can produce stable patterns of population activity called attractors[12]. Noise (either externally[14,45] or internally generated[11,15]) may destabilize those states, driving the emergence of temporal dynamics based on the progression through metastable states. Network models with clustered architecture provide a parsimonious explanation for the state sequences that have been observed ubiquitously in alert mammalian cortex, during both task engagement[17,18,46,47] and inter-trial periods.[11,39,40] In addition, this type of models accounts for various physiological observations such as stimulus-induced reduction of trial-to-trial variability[11,14,15,48], neural dimensionality[16], and firing rate multistability[11] (see also[49,50]). In particular, models with metastable attractors have been used to explain the state sequences observed in rodent gustatory cortex during taste processing and decision making[11,45,51].

In this work, we propose that clustered networks have the ability to modulate coding latency, and demonstrate one specific mechanism for modulation that can underlie the phenomenon of general expectation. Changes in the depth of attractor wells, induced by a non-stimulus specific anticipatory cue (which in turn may depend on the activation of top-down and neuromodulatory afferents[6,52]), can accelerate or slow down network dynamics. The acceleration resulting from shallower wells leads to a reshaping of ongoing activity and to a quicker recruitment of states coding for sensory information. To our knowledge, the link between generic anticipatory cues, network metastability, and coding speed as presented here is novel and represents the main innovation of our work.

**Functional role of heterogeneity in cue responses**

As stated in the previous section, the presence of clusters is a necessary ingredient to obtain a faster latency of coding. Here we discuss the second necessary ingredient, i.e., the presence of heterogeneous neural responses to the anticipatory cue (Fig. 1b).

Responses to anticipatory cues have been extensively studied in cortical and subcortical areas in alert rodents[6,10,53,54]. Cues evoke heterogeneous patterns of activity, either exciting or inhibiting single neurons. The proportion of cue responses and their heterogeneity develops with associative learning,[10,54] suggesting a fundamental function of these patterns. In the generic expectation paradigm considered here, the anticipatory cue does not convey any information about the identity of the forthcoming tastant, rather it just signals the availability of a stimulus. Experimental evidence suggests that the cue may induce a state of arousal, which was previously described as "priming" the sensory cortex[6,55,56]. Here, we propose an underlying mechanism in which the cue is responsible for acceleration of coding by increasing the spatial variance of pre-stimulus activity. In turn, this modulates the shape of the neuronal current-to-rate transfer function and thus lowers the effective energy barriers between metastable configurations.

We note that the presence of both excited and inhibited cue responses poses a challenge to simple models of neuromodulation. The presence of cue-evoked suppression of firing[10] suggests that cues do not improve coding by simply increasing the excitability of cortical areas. Additional mechanisms and complex patterns of connectivity may be required to explain the suppression effects induced by the cue. However, here we provide a parsimonious explanation of how heterogeneous responses can improve coding without postulating any specific pattern of connectivity other than i) random projections from thalamic and anticipatory cue afferents and ii) the clustered organization of the intra-cortical circuitry. Notice that the latter contains wide distributions of synaptic weights and can be understood as the consequence of Hebb-like re-organization of the circuitry during training[57,58].

It is also worth noting that our model incorporates excited and inhibited cue responses in such a manner to affect only the spatial variance of the activity across neurons, while leaving the mean input to the network unaffected. As a result, the anticipatory cue leaves average firing rates unchanged in the clustered network (Fig. S10), and only modulates the network temporal dynamics. Our model thus provides a mechanism whereby increasing the spatial variance of top-down inputs has, paradoxically, a beneficial effect on sensory coding.

**Specificity of the anticipatory mechanism**

Our model of anticipation relies on gain reduction in clustered excitatory neurons due to a larger spatial variance of the afferent currents. This model is robust to variations in parameters and architecture (Fig. S1, S4-S6). A priori, this effect might be achieved through different means, such as: increasing the strength of feedforward couplings; decreasing the strength of recurrent couplings; or modulating background synaptic inputs[59]. However, when scoring those models on the criteria of coding anticipation and heterogeneous cue responses, we found that they failed to simultaneously match both criteria, although for some range of parameters they could reproduce either one (see Fig. S8-9 and Supplementary Results for a detailed analysis). Thus, we concluded that only the main mechanism proposed here (Fig. 1a) captures the plurality of experimental observations pertaining anticipatory activity in a robust and biologically plausible way.

**Cortical timescales, state transitions, and cognitive function**

In populations of spiking neurons, a clustered architecture can generate reverberating activity and sequences of metastable states. Transitions from state to state can be typically caused by external inputs[11,15]. For instance, in frontal cortices, sequences of states are related to specific epochs within a task, with transitions evoked by behavioral events[18,19,22]. In sensory cortex, progressions through state sequences can be triggered by sensory stimuli and reflect the dynamics of sensory processing[23,46]. Importantly, state sequences have been observed also in the absence of any external stimulation, promoted by intrinsic fluctuations in neural activity[11,41]. However, the potential functional role, if any, of this type of ongoing activity has remained unexplored.

Recent work has started to uncover the link between ensemble dynamics and sensory and cognitive processes. State transitions in various cortical areas have been linked to decision making[45,60], choice representation[22], rule-switching behavior[24], and the level of task difficulty[23]. However, no theoretical or mechanistic explanations have been given for these phenomena.

Here we provide a mechanistic link between state sequences and expectation, by showing that intrinsically generated sequences can be accelerated, or slowed down, thus affecting sensory coding. Moreover, we show that the interaction between external stimuli and intrinsic dynamics does not result in the simple triggering of state transitions, but rather in the modulation of the intrinsic transition probabilities. The modulation of intrinsic activity can dial the duration of states, producing either shorter or longer timescales. A shorter timescale leads to faster state sequences and coding anticipation after stimulus presentation (Fig. 1 and 4). Other external perturbations may induce different effects: for example, recruiting the network's inhibitory population slows down the timescale, leading to a slower coding (Fig. S7).

The interplay between intrinsic dynamics and anticipatory influences presented here is a novel mechanism for generating diverse timescales, and may have rich computational consequences. We demonstrated its function in increasing coding speed, but its role in mediating cognition is likely to be broader and calls for further explorations.

## Author Contributions

## Acknowledgements

This work was supported by a National Institute of Deafness and Other Communication Disorders Grant K25-DC013557 (LM), by the Swartz Foundation Award 66438 (LM), by National Institute of Deafness and Other Communication Disorders Grants NIDCD R01DC012543 and R01DC015234 (AF), and partly by a National Science Foundation Grant IIS-1161852 (GLC). The authors would like to thank Drs. S. Fusi, A. Maffei, G. Mongillo, and C. van Vreeswijk for useful discussions.

## Competing Financial Interests

The authors declare no competing financial interests.

# Figures

## Fig. 1

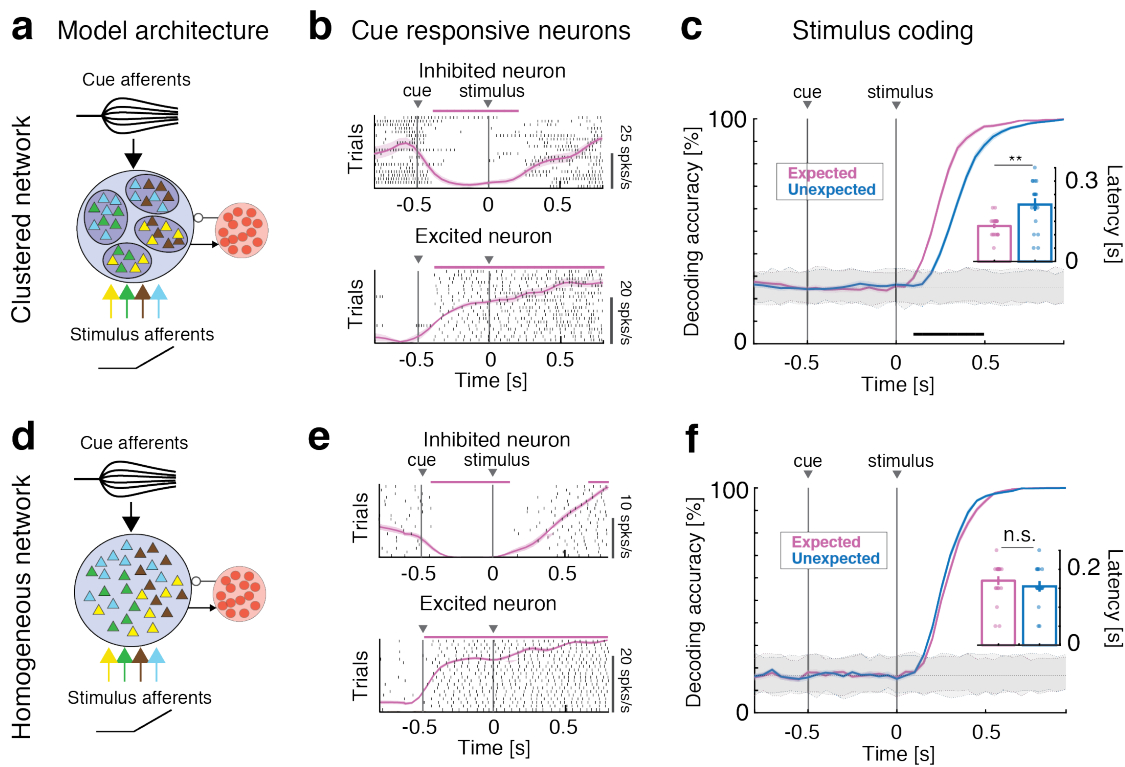

**Fig. 1**: **Anticipatory activity requires a clustered network architecture.** Effects of anticipatory cue on stimulus coding in the clustered (**a-c**) and homogeneous (**d-f**) network. **a:** Schematics of the clustered network architecture and stimulation paradigm. A recurrent network of inhibitory (red circles) and excitatory neurons (triangles) arranged in clusters (ellipsoids) with stronger intra-cluster recurrent connectivity. The network receives bottom-up sensory stimuli targeting random, overlapping subsets of clusters (selectivity to 4 stimuli is color-coded), and one top-down anticipatory cue inducing a spatial variance in the cue afferent currents to excitatory neurons. **b**: Representative single neuron responses to cue and one stimulus in expected trials in the clustered network of a). Black tick marks represent spike times (rasters), with PSTH (mean±s.e.m.) overlaid in pink. Activity marked by horizontal bars was significantly different from baseline (pre-cue activity) and could either be excited (top panel) or inhibited (bottom) by the cue. **c**: Time course of cross-validated stimulus-decoding accuracy in the clustered network. Decoding accuracy increases faster during expected (pink) than unexpected (blue) trials in clustered networks (curves and color-shaded areas represent mean±s.e.m. across four tastes in 20 simulated sessions; color-dotted

lines around gray shaded areas represent 95% C.I. from shuffled datasets). A separate classifier was used for each time bin, and decoding accuracy was assessed via a cross-validation procedure, yielding a confusion matrix whose diagonal represents the accuracy of classification for each of four tastes in that time bin (see text and Fig. S3 for details). *Inset*: aggregate analysis across simulated sessions of the onset times of significant decoding in expected (pink) vs. unexpected trials (blue). **d**: Schematics of the homogenous network architecture. Sensory stimuli modeled as in a). **e**: Representative single neuron responses to cue and one stimulus in expected trials in the homogeneous network of d), same conventions as in b). **f**: Cross-validated decoding accuracy in the homogeneous network (same analysis as in c). The latency of significant decoding in expected vs. unexpected trials is not significantly different. *Inset*: aggregate analysis of onset times of significant decoding (same as inset of c). Panels **b**, **c**, **e**, **f**: pink and black horizontal bars, *p < 0.05*, *t*-test with multiple-bin Bonferroni correction. Panel **c**: ** = *p < 0.01*, *t*-test. Panel **f**: *n.s.*: non-significant.

**Fig. 2**

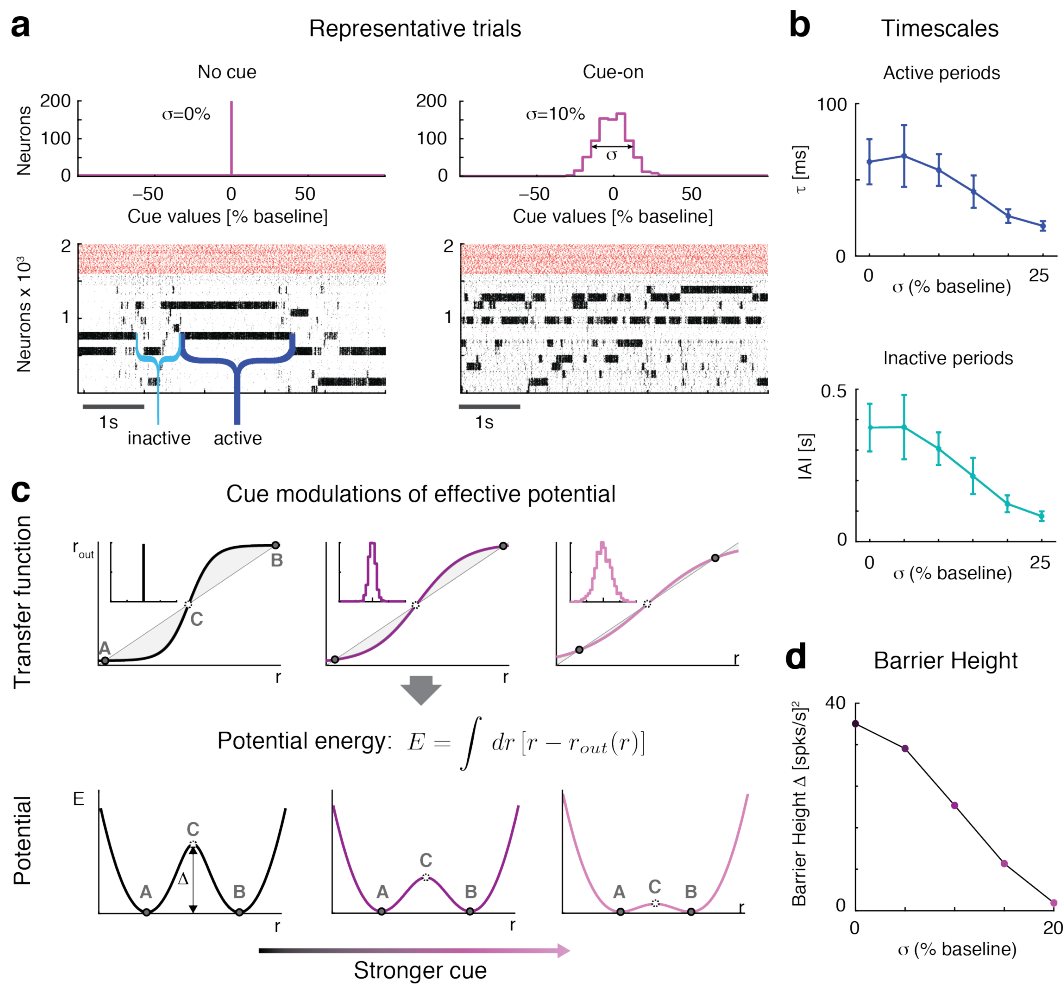

**Fig. 2**: **Anticipatory cue speeds up network dynamics**. **a**: Raster plots of the clustered network activity in the absence (left) and in the presence of the anticipatory cue (right), with no stimulus presentation in either case. The dynamics of cluster activation and deactivation accelerated proportionally to the increase in afferent currents' variance $\sigma^2$ induced by the cue. *Top panels*: distribution of cue peak values across excitatory neurons: left, no cue; right, distribution with S.D. $\sigma = 10\%$ in units of baseline current. *Bottom panels*: raster plots of representative trials in each condition (black: excitatory neurons, arranged according to cluster membership; red: inhibitory neurons). **b**: The average cluster activation lifetime (left) and inter-activation interval (right) significantly decrease when increasing $\sigma$. **c**: Schematics of the effect of the anticipatory cue on network dynamics. *Top row*: the increase in the spatial variance of cue afferent currents (*insets*: left: no cue; stronger cues towards the right) flattens the "effective *f-I* curve"

(sigmoidal curve) around the diagonal representing the identity line (straight line). The case for a simplified two-cluster network is depicted (see text). States A and B correspond to stable configurations with only one cluster active; state C corresponds to an unstable configuration with 2 clusters active. *Bottom row*: shape of the effective potential energy corresponding to the *f-I* curves shown in the top row. The effective potential energy is defined as the area between the identity line and the effective *f-I* curve (shaded areas in top row; see formula). The *f-I* curve flattening due to the anticipatory cue shrinks the height $\Delta$ of the effective energy barrier, making cluster transitions more likely and hence more frequent. **d**: Effect of the anticipatory cue (in units of the baseline current) on the height of the effective energy barrier $\Delta$, calculated via mean field theory in a reduced two-cluster network of LIF neurons (see Methods).

**Fig. 3**

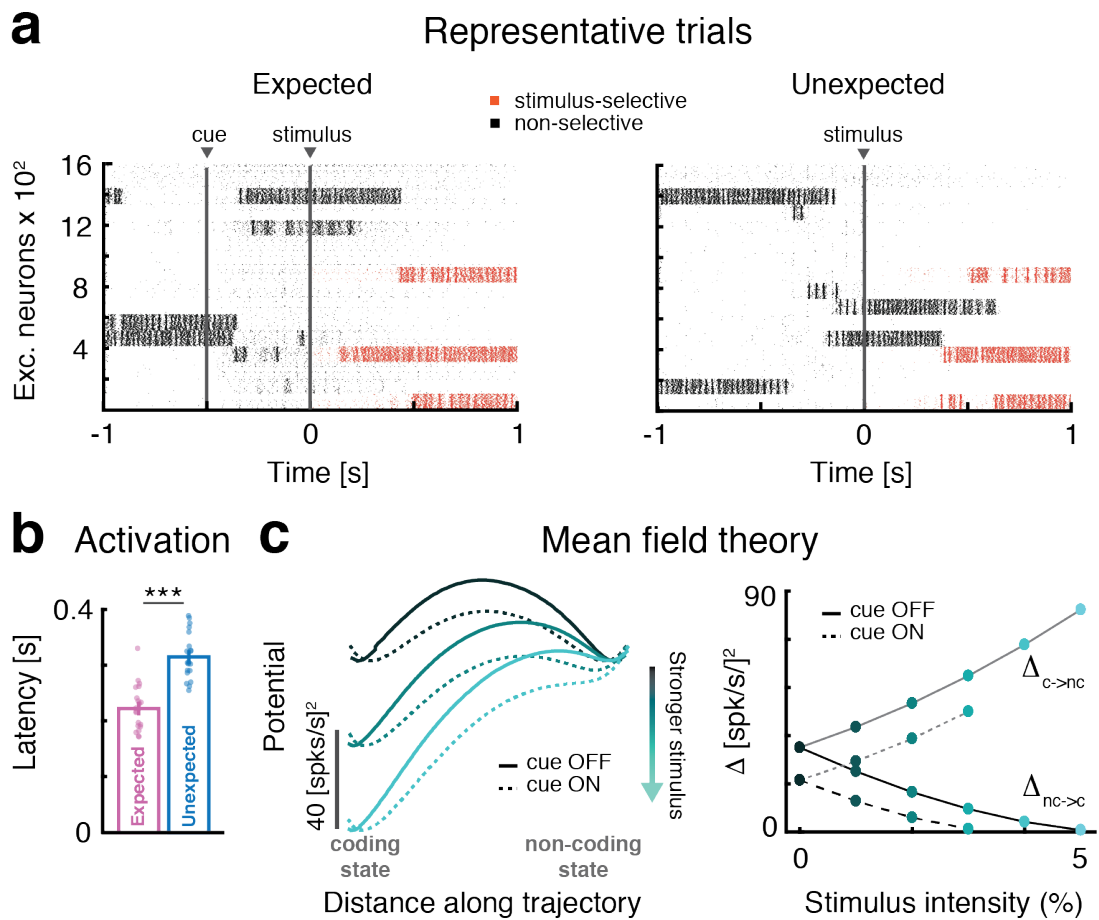

**Fig. 3**: **Anticipatory cue induces faster onset of stimulus-coding states. a:** Raster plots of representative trials in the expected (left) and unexpected (right) conditions in response to the same stimulus at *t=0*. Stimulus-selective clusters (red tick marks, spikes) activate earlier than non-selective clusters (black tick marks, spikes) in response to the stimulus when the cue precedes the stimulus. **b**: Comparison of activation latency of selective clusters after stimulus presentation during expected (pink) and unexpected (blue) trials (mean±s.e.m. across 20 simulated sessions). Latency in expected trials is significantly reduced. **c**: The effective energy landscape and the modulation induced by stimulus and anticipatory cue on two-clustered networks, computed via mean field theory (see Methods). *Left panel*: after stimulus presentation the stimulus-coding state (left well in left panel) is more likely to occur than the non-coding state (right well). *Right panel*: barrier heights as a function of stimulus strength in expected (`cue ON') and unexpected trials (`cue OFF'). Stronger stimuli (lighter shades of cyan) decrease the barrier height $\Delta$ separating the non-coding and the coding state. In expected trials

(dashed lines), the barrier Δ is smaller than in unexpected ones (full lines), leading to a faster transition probability from non-coding to coding states compared to unexpected trials (for stimulus ≥ 4% the barrier vanishes leaving just the coding state). Panel **b**: *** = *p < 0.001*, *t*-test.

**Fig. 4**

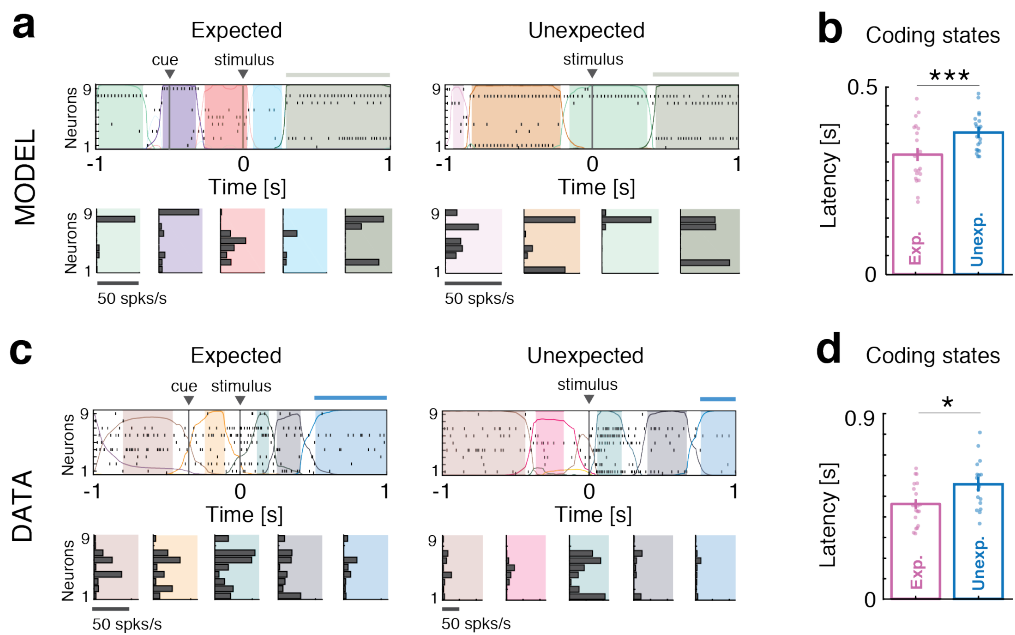

**Fig. 4**: **Anticipation of coding states: model vs. data**. **a:** Representative trials from one ensemble of 9 simultaneously recorded neurons from clustered network simulations during expected (left) and unexpected (right) conditions. *Top panels*: spike rasters with latent states extracted via a HMM analysis (colored curves represent time course of state probabilities; colored areas indicate intervals where the probability of a state exceeds 80%; thick horizontal bars atop the rasters mark the presence of a stimulus-coding state). *Bottom panels*: Firing rate vectors for each latent state shown in the corresponding top panel. **b**: Latency of stimulus-coding states in expected (pink) vs. unexpected (blue) trials (mean±s.e.m. across 20 simulated datasets). Faster coding latency during expected trials is observed compared to unexpected trials. **c-d:** Same as a)-b) for the empirical datasets. *= p < 0.05, ***= p < 0.001*, t-test.

# Methods

*Experimental dataset.* The experimental data come from a previously published dataset Ref.[1] (for details, see Supplementary Methods). Experimental procedures were approved by the Institutional Animal Care and Use Committee of Stony Brook University and complied with university, state, and federal regulations on the care and use of laboratory animals.

*Ensemble states detection.* A Hidden Markov Model (HMM) analysis was used to detect ensemble states in both the empirical data and model simulations. Here, we give a brief description of the method used and we refer the reader to Refs. [2-5] for more detailed information.

The HMM assumes that an ensemble of $N$ simultaneously recorded neurons is in one of $M$ hidden states at each given time bin. States are firing rate vectors $r_i(m)$, where $i = 1,\ldots,N$ is the neuron index and $m = 1,\ldots,M$ identifies the state. In each state, neurons were assumed to discharge as stationary Poisson processes (Poisson-HMM) conditional on the state's firing rates. Trials were segmented in 2 ms bins, and the value of either 1 (spike) or 0 (no spike) was assigned to each bin for each given neuron (Bernoulli approximation for short time bins); if more than one neuron fired in a given bin (a rare event), a single spike was randomly assigned to one of the firing neurons. A single HMM was used to fit all trials in each recording session, resulting in the emission probabilities $r_i(m)$ and in a set of transition probabilities between the states. Emission and transition probabilities were calculated with the Baum-Welch algorithm[6] with a fixed number of hidden states $M$, yielding a maximum likelihood estimate of the parameters given the observed spike trains. Since the model log-likelihood $LL$ increases with $M$, we repeated the HMM fits for increasing values of $M$ until we hit a minimum of the Bayesian Information Criterion (BIC, see below and Ref.[6]). For each $M$, the $LL$ used in the BIC was the sum over 10 independent HMM fits with random initial guesses for emission and transition probabilities. This step was needed since the Baum-Welch algorithm only guarantees reaching a local rather than global maximum of the likelihood. The model with the lowest BIC (having $M^*$ states) was selected as the winning model, where $BIC = -2LL + [M(M-1) + MN] \cdot \ln T$, $T$ being the number of observations in each session (= number of trials × number of bins per trials). Finally, the winning HMM model was used to "decode" the states from the data according to their posterior probability given the data. During decoding, only those states with probability exceeding 80% in at least 25 consecutive 2ms-bins were retained (henceforth denoted simply as "states")[3,5]. This procedure eliminates states that appear only very transiently and with low probability, also reducing the chance of overfitting. A median of 6 states per ensemble was found, varying from 3 to 9 across ensembles.

*Coding states.* In each condition (i.e., expected vs. unexpected), the frequencies of occurrence of a given state across taste stimuli were compared with a test of proportions (chi-square, p<0.001 with Bonferroni correction to account for multiple states). When a significant difference was found across stimuli, a post-hoc Marascuilo test was performed[7]. A state whose frequency of occurrence was significantly higher in the presence of one taste stimulus compared to all other tastes was deemed a 'coding state' for that stimulus (Fig. 4).

*Spiking network model.* We modeled the local neural circuit as a recurrent network of $N$ leaky-integrate-and-fire (LIF) neurons, with a fraction $n_E = 80\%$ of excitatory (E) and $n_I = 20\%$ of inhibitory (I) neurons.[8] Connectivity was random with probability $p_{EE} = 0.2$ for E to E connections and $p_{EI} = p_{IE} = p_{II} = 0.5$ otherwise. Synaptic weights $J_{ij}$ from pre-synaptic neuron $j \in E, I$ to post-synaptic neuron $i \in E, I$ scaled as $J_{ij} = j_{ij}/\sqrt{N}$, with $j_{ij}$ drawn from normal distributions with mean $j_{\alpha\beta}$ (for $\alpha, \beta = E, I$) and 1% SD. Networks of different architectures were considered: *i)* networks with segregated clusters (referred to as "clustered network," parameters as in Tables 1 and 2); *ii)* networks with overlapping clusters (see Suppl. Table S2 and Suppl. Methods for details), *iii)* homogeneous networks (parameters as in Table 1). In the clustered network, E neurons were arranged in $Q$ clusters with $N_c = 100$ neurons per clusters on average (1% SD), the remaining fraction $n_{bg}$ of E neurons belonging to an unstructured "background" population. In the clustered network, neurons belonging to the same cluster had intra-cluster synaptic weights potentiated by a factor $J_+$; synaptic weights between neurons belonging to different clusters were depressed by a factor $J_- = 1 - \gamma f(J_+ - 1) < 1$ with $\gamma = 0.5$; $f = (1 - n_{bg})/Q$ is the average number of neurons in each cluster.[8] When changing the network size $N$, all synaptic weights $J_{ij}$ were scaled by $\sqrt{N}$, the intra-cluster potentiation values were $J_+$=5, 10, 20, 30, 40 for $N = 1, 2, 4, 6, 8 \times 10^3$ neurons, respectively, and cluster size remained unchanged (see also Table 1); all other parameters were kept fixed. In the homogeneous network, there were no clusters ($J_+ = J_- = 1$).

*Model neuron dynamics.* Below threshold the LIF neuron membrane potential evolved in time as

$$\frac{d}{dt}V = -\frac{V}{\tau_m} + I_{rec} + I_{ext},$$

where $\tau_m$ is the membrane time constant and the input currents $I$ are a sum of a recurrent contribution $I_{rec}$ coming from the other network neurons and an external current $I_{ext} = I_0 + I_{stim} + I_{cue}$ (units of Volt/s). Here, $I_0$ is a constant term representing input from other brain areas; $I_{stim}$ and $I_{cue}$ represent the incoming stimuli and cue, respectively (see *Stimulation protocols* below). When $V$ hits threshold $V_{thr}$, a spike is emitted and $V$ is then clamped to the rest value $V_{reset}$ for a refractory period $\tau_{ref}$. Thresholds were chosen so that the homogeneous network neurons fired at rates $r_E = 5$ spks/s and $r_I = 7$ spks/s. The recurrent contribution to the postsynaptic current to the $i$-th neuron was a low-pass filter of the incoming spike trains

$$\tau_{syn} \frac{d}{dt} I_{rec} = -I_{rec} + \sum_{j=1}^{N} J_{ij} \sum_{k} \delta(t - t_k^j),$$

where $\tau_{syn}$ is the synaptic time constant; $J_{ij}$ is the recurrent synaptic weights from presynaptic neuron $j$ to postsynaptic neuron $i$, and $t_k^j$ is the $k$-th spike time from the $j$-th presynaptic neuron. The constant external current was $I_0 = N_{ext} p_{i0} J_{i0} v_{ext}$, with $N_{ext} = n_E N$, $p_{i0} = 0.2$, $J_{i0} = j_{i0}/\sqrt{N}$ with $j_{E0}$ for excitatory and $j_{I0}$ for inhibitory neurons (see Table 1), and $r_{ext} = 7$ spks/s. For a detailed mean field theory analysis of the clustered network and a comparison between simulations and mean field theory during ongoing and stimulus-evoked periods we refer the reader to the Suppl. Methods and Refs.[5,9].

*Stimulation protocols.* Stimuli were modeled as time-varying stimulus afferent currents targeting 50% of neurons in stimulus-selective clusters $I_{stim}(t) = I_0 \cdot r_{stim}(t)$, where $r_{stim}(t)$ was expressed as a fraction of the baseline external current $I_0$. Each cluster had a 50% probability of being selective to a given stimulus, thus different stimuli targeted overlapping sets of clusters. The anticipatory cue, targeting a random 50% subset of E neurons, was modeled as a double exponential with rise and decay times of 0.2 s and 1 s, respectively, unless otherwise specified; its peak value for each selective neuron was sampled from a normal distribution with zero mean and standard deviation $\sigma$ (expressed as fraction of the baseline current $I_0$; $\sigma$=20% unless otherwise specified). The cue did not change the mean afferent current but only its spatial (quenched) variance across neurons.

In both the unexpected and the unexpected conditions, stimulus onset at $t = 0$ was followed by a linear increase $r_{stim}(t)$ in the stimulus afferent current to stimulus-selective neurons reaching a value $r_{max}$ at $t = 1$ s ($r_{max} = 20\%$, unless otherwise specified). In the expected condition, stimuli were preceded by the anticipatory cue $r_{cue}(t)$ with onset at $t = -0.5$s before stimulus presentation.

*Network simulations.* All data analyses, model simulations and mean field theory calculations were performed using custom software written in MATLAB (MathWorks), and C. Simulations comprised 20 realizations of each network (each one representing a different experimental session), with 20 trials per stimulus in each of the 2 conditions (unexpected and expected); or 40 trials per session in the condition with "cue-on" and no stimuli (Fig. 2). Dynamical equations for the LIF neurons were integrated with the Euler method with 0.1 ms step.

*Mean field theory*. Mean field theory was used in a simplified network with 2 excitatory clusters (parameters as in Table 2) using the population density approach[10-12]: the input to each neuron was completely characterized by the infinitesimal mean $\mu_\alpha$ and variance $\sigma_\alpha^2$ of the post-synaptic current (see Sec. 2.3 of Supplementary Methods for their expressions). The network fixed points satisfied the $Q + 2$ self-consistent mean field equations[8]

$$r_\alpha = F_\alpha(\mu_\alpha(\boldsymbol{r}), \sigma_\alpha^2(\boldsymbol{r})), \tag{1}$$

where $\boldsymbol{r} = [r_1, \ldots, r_Q, r_E^{bg}, r_I]$ is the population firing rate vector (boldface represents vectors). $F_\alpha$ is the current-to-rate function for population $\alpha$, which varied depending on the population and the condition. In the absence of the anticipatory cue, the LIF current-to-rate function $F_\alpha^0$ was used

$$F_\alpha^0(\mu_\alpha, \sigma_\alpha) = \left( \tau_{ref} + \tau_{m,\alpha} \sqrt{\pi} \int_{H_{eff,\alpha}}^{\Theta_{eff,\alpha}} e^{u^2} [1 + \mathrm{erf}(u)] \right)^{-1},$$

where $\Theta_{eff,\alpha} = \frac{V_{thr,\alpha} - \mu_\alpha}{\sigma_\alpha} + ak_\alpha$, $H_{eff,\alpha} = \frac{V_{reset,\alpha} - \mu_\alpha}{\sigma_\alpha} + ak_\alpha$. Here, $k_\alpha = \sqrt{\tau_{syn,\alpha}/\tau_{m,\alpha}}$, $a = \frac{|\zeta(1/2)|}{\sqrt{2}} \sim 1.03$.[13,14] In the presence of the anticipatory cue, a modified current-to-rate function $F_\alpha^{cue}$ was used to capture the cue-induced Gaussian noise in the cue afferent currents to the cue-selective populations ($\alpha = 1, \ldots, Q$):

$$F_\alpha^{cue}(\mu_\alpha, \sigma_\alpha) = \int Dz\, F_\alpha^0(\mu_\alpha + z\sigma\mu_{ext}, \sigma_\alpha),$$

where $Dz = dz \exp(-\frac{z^2}{2})/\sqrt{2\pi}$ is the Gaussian measure with zero mean and unit variance, $\mu_{ext} = I_0$ is the baseline afferent current and $\sigma$ is the anticipatory cue's SD as fraction of $\mu_{ext}$ (Fig. 3d; in Fig. 3c, for illustration purposes we used $F_E^0 = 0.5(1 + \tanh)$. Fixed points $\boldsymbol{r}^*$ of equation (1) were found with Newton's method; the fixed points were stable (attractors) when the stability matrix

$$S_{\alpha\beta} = \frac{1}{\tau_{syn,\alpha}} \left( \frac{\partial F_\alpha(\mu_\alpha(\boldsymbol{r}), \sigma_\alpha^2(\boldsymbol{r}))}{\partial r_\beta} - \frac{\partial F_\alpha(\mu_\alpha(\boldsymbol{r}), \sigma_\alpha^2(\boldsymbol{r}))}{\partial \sigma_\alpha^2} \frac{\partial \sigma_\alpha^2}{\partial r_\beta} - \delta_{\alpha\beta} \right), \tag{2}$$

evaluated at $\boldsymbol{r}^*$ was negative definite. Stability was defined with respect to an approximate linearized dynamics of the mean $m_\alpha$ and SD $s_\alpha$ of the input currents[15]

$$\tau_{syn,\alpha}\frac{dm_\alpha}{dt} = -m_\alpha + \mu_\alpha(\mathbf{r}),$$
$$\frac{\tau_{syn,\alpha}}{2}\frac{ds_\alpha^2}{dt} = -s_\alpha^2 + \sigma_\alpha^2(\mathbf{r}), \quad (3)$$
$$r_\alpha(t) = F_\alpha(m_\alpha(\mathbf{r}), s_\alpha^2(\mathbf{r})),$$

where $\mu_\alpha$ and $\sigma_\alpha^2$ are the stationary values given in the Suppl. Methods.

*Effective mean field theory for the reduced network.* The mean field equations (1) for the $P=Q+2$ populations may be reduced to a set of effective equations governing the dynamics of a smaller subset of $q<P$ of populations, henceforth referred to as populations *in focus*[16]. The reduction is achieved by integrating out the remaining $P-q$ *out-of-focus* populations. This procedure was used to estimate the energy barrier separating the two network attractors in Fig. 2d and Fig. 3c. Given a fixed set of values $\tilde{\mathbf{r}} = [\tilde{r}_1, \dots, \tilde{r}_q]$ for the in-focus populations, one solves the mean field equations for $P-q$ out-of-focus populations

$$r_\beta(\tilde{r}_1, \dots, \tilde{r}_q) = F_\beta[\mu_\beta(\tilde{r}_1, \dots, \tilde{r}_q, r_{q+1}, \dots, r_P), \sigma_\beta^2(\tilde{r}_1, \dots, \tilde{r}_q, r_{q+1}, \dots, r_P)],$$

for $\beta = q+1, \dots, P$ to obtain the stable fixed point $\mathbf{r}'(\tilde{\mathbf{r}}) \doteq [r_{q+1}(\tilde{\mathbf{r}}), \dots, r_P(\tilde{\mathbf{r}})]$ of the out-of-focus populations as functions of the in-focus firing rates $\tilde{\mathbf{r}}$. Stability of the solution $\mathbf{r}'(\tilde{\mathbf{r}})$ is computed with respect to the stability matrix (2) of the reduced system of $P-q$ out-of-focus populations. Substituting the values $\mathbf{r}'(\tilde{\mathbf{r}})$ into the fixed-point equations for the $q$ populations in focus yields a new set of equations relating input rates $\tilde{\mathbf{r}}$ to "output" rates $\mathbf{r}^{out}$:

$$r_\alpha^{out}(\tilde{\mathbf{r}}) = F_\alpha[\mu_\alpha(\tilde{\mathbf{r}}, \mathbf{r}'(\tilde{\mathbf{r}})), \sigma_\alpha^2(\tilde{\mathbf{r}}, \mathbf{r}'(\tilde{\mathbf{r}}))],$$

for $\alpha = 1, \dots, q$. The input $\tilde{\mathbf{r}}$ and output $\mathbf{r}^{out}$ firing rates of the in-focus populations will be different, except at a fixed point of the full system where they coincide. The correspondence between input and output rates of in-focus populations defines the effective current-to-rate transfer functions

$$r_\alpha^{out}(\tilde{\mathbf{r}}) = F_\alpha^{eff}[\mu_\alpha(\tilde{\mathbf{r}}), \sigma_\alpha^2(\tilde{\mathbf{r}})], \quad (4)$$

for $\alpha = 1, \dots, q$ in-focus populations at the point $\tilde{\mathbf{r}}$. The fixed points $r_\alpha^{out}(\tilde{\mathbf{r}}^*) = \tilde{r}_\alpha^*$ of the in-focus equations (4) are fixed points of the entire system. It may occur, in general, that the out-of-focus populations attain multiple attractors for a given value of $\tilde{\mathbf{r}}$, in which case the set of effective transfer functions $F_\alpha^{eff}$ is labeled by the chosen attractor; in our analysis of the two-clustered network, only one attractor was present for a given value of $\tilde{\mathbf{r}}$.

*Energy potential.* In a network with $Q=2$ clusters, one can integrate out all populations (out-of-focus) except one (in-focus) to obtain the effective transfer functions for the in-focus population representing a single cluster, with firing rate $\tilde{r}$ (equation (4) for $q=1$). Network dynamics can be visualized on a one-dimensional curve, where it is well approximated by the first-order dynamics (see ref. [16] for details):

$$\tau_{syn,\alpha} \frac{d\tilde{r}}{dt} = -\tilde{r} + r^{out}(\tilde{r}).$$

These dynamics can be expressed in terms of an effective energy function $E(\tilde{r})$ as

$$\tau_{syn,\alpha} \frac{d\tilde{r}}{dt} = -\frac{\partial E(\tilde{r})}{\partial \tilde{r}},$$

so that the dynamics can be understood as a motion in an effective potential energy landscape, as if driven by an effective force $-\frac{\partial E(\tilde{r})}{\partial \tilde{r}} = -(\tilde{r} - r^{out}(\tilde{r}))$. The minima of the energy with respect to $\tilde{r}$ are the stable fixed points of the effective 1-dimensional dynamics, while its maxima represent the effective energy barriers between two minima, as illustrated in Fig. 2c. The one-cluster network has 3 fixed points, two stable attractors ('A' and 'B' in Fig. 2c) and a saddle point ('C'). We estimated the height $\Delta$ of the potential energy barrier on the trajectory from A to B through C as minus the integral of the force from the first attractor A to C:

$$\Delta = \int_A^C (\tilde{r} - r^{out}(\tilde{r})) d\tilde{r},$$

which represents the area between the identity line ($y = \tilde{r}$) and the effective transfer function ($y = r^{out}(\tilde{r})$) (see Fig. 2c). In the finite network, where the dynamics comprise stochastic transitions among the states, switching between A and B would occur with a frequency that depends on the effective energy barrier $\Delta$, as explained in the main text.

*Population decoding.* The amount of stimulus-related information carried by spike trains was assessed through a decoding analysis[17] (see Fig. S3a for illustration). A multiclass classifier was constructed from $Q$ neurons sampled from the population (one neuron from each of the $Q$ clusters for clustered networks, or $Q$ random excitatory neurons for homogeneous networks). Spike counts from all trials of $n_{stim}$ taste stimuli in each condition (expected vs. unexpected) were split into training and test sets for cross-validation. A "template" was created for the population PSTH for each stimulus, condition and time bin (200 ms, sliding over in 50 ms steps) in the training set. The PSTH contained the trial-averaged spike counts of each neuron in each bin (the same number of trials across stimuli and conditions were used). Population spike counts for each test trial were classified according to the smallest Euclidean distance from the templates across 10 training sets ('bagging' or bootstrap aggregating procedure[18]). Specifically, from each training set $L$, we created bootstrapped training sets $L_b$, for $b = 1,..,B=10$, by sampling with replacement from $L$. In each bin, each test trial was then classified $B$ times using the $B$ classifiers, obtaining $B$ different "votes", and the most frequent vote was chosen as the bagged classification of the test trial. Cross-validated decoding accuracy in a given bin was defined as the fraction of correctly classified test trials in that bin.

Significance of decoding accuracy was established via a permutation test: 1000 shuffled datasets were created by randomly permuting stimulus labels among trials, and a 'shuffled distribution' of 1000 decoding accuracies was obtained. In each bin, decoding accuracy of the original dataset was deemed significant if it exceeded the upper bound, $\alpha_{0.05}$, of the 95% confidence interval of the shuffled accuracy distribution in that bin (this included a Bonferroni correction for multiple bins, so that $\alpha_{0.05} = 1 - 0.05/N_b$, with $N_b$ the number of bins). Decoding latency (insets in Figs. 1c and 1f) was estimated as the earliest bin with significant decoding accuracy.

*Cluster dynamics.* To analyze the dynamics of neural clusters (lifetime, inter-activation interval, and latency; see Figs. 2 and 3), cluster spike count vectors $r_i$ (for $i = 1, \ldots, Q$) in 5 ms bins were obtained by averaging spike counts of neurons belonging to a given cluster. A cluster was deemed active if its firing rate exceeded 10 spks/s. This threshold was chosen so as to lie between the inactive and active clusters' firing rates, which were obtained from a mean field solution of the network[5].

*Code and data availability statement.* Experimental datasets are available from the corresponding authors upon request. All data analysis and network simulation scripts are available from the authors upon request.

| Symbol | Description | Value |
|---|---|---|
| $j_{EE}$ | Mean E→E synaptic weight $\times \sqrt{N}$ | 1.1 mV |
| $j_{EI}$ | Mean I→E synaptic weight $\times \sqrt{N}$ | 5.0 mV |
| $j_{IE}$ | Mean E→I synaptic weight $\times \sqrt{N}$ | 1.4 mV |
| $j_{II}$ | Mean I→I synaptic weight $\times \sqrt{N}$ | 6.7 mV |
| $j_{E0}$ | Mean afferent synaptic weights to E neurons $\times \sqrt{N}$ | 5.8 mV |
| $j_{I0}$ | Mean afferent synaptic weights to I neurons $\times \sqrt{N}$ | 5.2 mV |
| $J_+$ | Potentiated intra-cluster E→E weights factor | see caption |
| $r_{ext}^E$ | Average afferent rate to E neurons (baseline) | 7 spks/s |
| $r_{ext}^I$ | Average afferent rate to I neurons (baseline) | 7 spks/s |
| $V_{thr}^E$ | E neuron threshold potential | 3.9 mV |
| $V_{thr}^I$ | I neuron threshold potential | 4.0 mV |
| $V_{reset}$ | E and I neurons reset potential | 0 mV |
| $\tau_m$ | E and I membrane time constant | 20 ms |
| $\tau_{ref}$ | Absolute refractory period | 5 ms |
| $\tau_{syn}$ | E and I synaptic time constant | 4 ms |
| $n_{bg}$ | Background neurons fraction | 10% |
| $N_c$ | Average cluster size | 100 neurons |

Table 1: Parameters for the clustered and homogeneous networks with *N* LIF neurons. In the clustered network, the intra-cluster potentiation parameter values were $J_+$= 5, 10, 20, 30, 40 for networks with $N = 1, 2, 4, 6, 8 \times 10^3$ neurons, respectively. In the homogeneous network, $J_+$=1.

| Symbol | Description | Value |
|---|---|---|
| $j_{EE}$ | Mean E→E synaptic weight $\times \sqrt{N}$ | 0.8 mV |
| $j_{EI}$ | Mean I→E synaptic weight $\times \sqrt{N}$ | 10.6 mV |
| $j_{IE}$ | Mean E→I synaptic weight $\times \sqrt{N}$ | 2.5 mV |
| $j_{II}$ | Mean I→I synaptic weight $\times \sqrt{N}$ | 9.7 mV |
| $j_{E0}$ | Mean afferent synaptic weights to E neurons $\times \sqrt{N}$ | 14.5 mV |
| $j_{I0}$ | Mean afferent synaptic weights to I neurons $\times \sqrt{N}$ | 12.9 mV |
| $J_+$ | Potentiated intra-cluster E→E weights factor | 9 |
| $V_{thr}^E$ | E neuron threshold potential | 4.6 mV |
| $V_{thr}^I$ | I neuron threshold potential | 8.7 mV |
| $n_{bg}$ | Background neurons fraction | 65% |

Table 2: Parameters for the simplified 2-cluster network with N=800 LIF neurons (the remaining parameters were as in Table 1).

# Expectation-induced modulation of metastable activity underlies faster coding of sensory stimuli

**Supplementary Material**

---

**Luca Mazzucato**[1,2] **Giancarlo La Camera\***[1,3] **and Alfredo Fontanini\***[1,3]

[1]*Department of Neurobiology and Behavior, State University of New York at Stony Brook, Stony Brook, NY 11794*

[2]*Departments of Biology and Mathematics and Institute of Neuroscience, University of Oregon, Eugene, OR 97403*

[3]*Graduate Program in Neuroscience, State University of New York at Stony Brook, Stony Brook, NY 11794*

# Supplementary Results

## 1.1 Cue responses in data vs. model

We compared the cue responses of the clustered network to those in the dataset published in Ref. [1]. The entity of the response was estimated by the ∆PSTH with the procedure described in the Suppl. Methods, Sec. 2.2. In the data [1], the PSTH modulation in cue-responsive neurons was consistent across modalities (excited responses peaked at $12 \pm 1.0$ spks/s, inhibited responses peaked at $-4.8 \pm 0.3$ spks/s; see Fig. S1a). In the model, modulation of cue responses depended on the cue-induced modulation of the spatial variance $\sigma^2$ of the afferent currents (expressed in percent of baseline in Fig. S1b). As shown in the figure, the PSTH modulation of cue-responsive neurons slightly increased with $\sigma$, however, over a wide range of parameters, was in quantitative agreement with the data shown in Fig. S1a.

In Fig. 1a–c of the main text we chose $\sigma = 20\%$, giving peak firing rate modulations of $11.1 \pm 0.8$ spks/s (excited) and $-6.2 \pm 0.3$ spks/s (inhibited). Single neurons' responses thus reflected responses observed in the empirical data (Fig. S1a and refs. [1, 2]). In the homogeneous network of Fig. 1d–f, single neurons' responses were comparable to the ones observed in the clustered network. With $\sigma = 20\%$, peak firing rate modulations were $11.2 \pm 1.2$ spks/s (excited) and $-3.9 \pm 0.2$ spks/s (inhibited). Both in the clustered and in the homogeneous networks, neurons could be nonresponsive to the cue (Fig. S2).

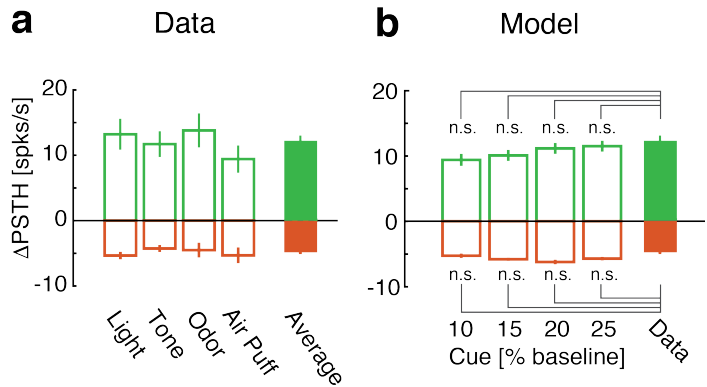

**Fig. S1**. **Cue responses in data and model**. Peak firing rate responses of cue-excited (green) and cue-inhibited (red) neurons. **a**: Data from different sensory modalities together with across-modalities average (rightmost bar), adapted from [1]. **b**: Clustered network model peak firing rate responses for different values of cue-induced spatial variance. Peak responses ($\Delta$PSTH, mean$\pm$s.e.m.) were computed in both panels as the difference between the activity post-cue and the activity pre-cue. All panels: $t$-test, n.s.=non-significant. "Data" in panel **b** is the same as "Average" in panel **a**.

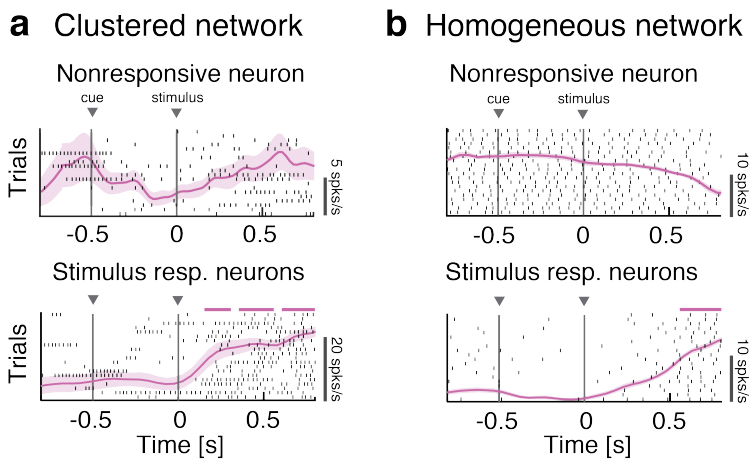

**Fig. S2**. **Model neurons with no cue response**. Raster plots of representative single neuron responses to cue and one stimulus in expected trials in the clustered (**a**) and homogeneous (**b**) networks (pink curve: PSTH (mean$\pm$s.e.m.). *Top row:* non-responsive neurons. *Bottom row:* neurons that are stimulus-responsive but not cue-responsive according to whether or not the PSTH was significantly different from baseline (pink horizontal bars: $p < 0.05$, $t$-test with multiple-bin correction).

### 1.2 Anticipation of stimulus decoding

In the main text we have shown that stimulus decoding is accelerated in the presence of the anticipatory cue (for example, in Fig. 1c). As a measure of performance we used the decoding accuracy averaged across all tastes. Here we show that similar results are obtained when considering the decoding performance for single tastes separately. In either case, the decoding procedure is described in Methods (Sec. *Population decoding*) and is summarized in Fig. S3a.

As shown in figure, decoding was performed using a multi-class classifier where each class corresponded to one of the four taste stimuli. A separate classifier was used for each time bin, and decoding performance in each bin was assessed via a cross-validation procedure (see Methods for details). This procedure yields a confusion matrix whose diagonal elements represent the accuracy of classification for each taste in that time bin (Fig. S3a). The performance for each time bin can be averaged across instances of the same taste (as in the left panel of Fig. S3b) or across all tastes simultaneously (right panel of Fig. S3b). In either case, significant decoding accuracy is anticipated in expected trials compared to unexpected trials.

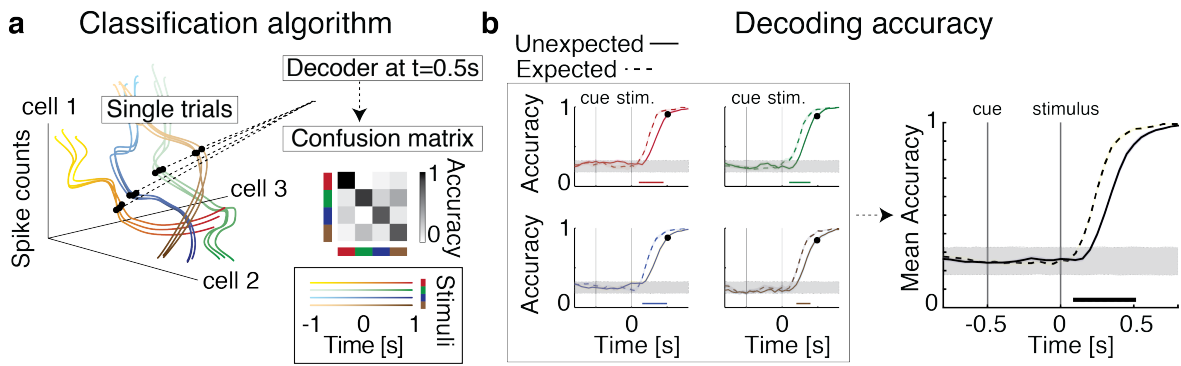

**Fig. S3**. **Multi-class classifier for stimulus decoding**. **a**: Schematics of the classification algorithm used for population decoding. The colored curves represent the temporal evolution of neural activity across $N$ simultaneously recorded neurons. The four colors label trajectories obtained with four different stimuli (tastants). Hues help visualize the time course within each trajectory. Each time bin of activity (black dots) was decoded using an independent classifier. Decoding performance was assessed via a cross-validation procedure, yielding a confusion matrix whose diagonal represents the accuracy of classification for each taste in that time bin. **b**: Time course of decoding accuracy in expected vs. unexpected conditions for each individual tastant (left) and for the across-taste average (right). The left panel demonstrates coding anticipation for each tastant separately. Color-coded horizontal bar represents significant difference between decoding accuracy in expected vs. unexpected trials, $p < 0.05$, $t$-test with multiple bin Bonferroni correction (notations as in Fig. 1c of the main text).

## 1.3 Robustness of anticipatory activity

To test the robustness of anticipatory activity in the clustered network, we systematically varied key parameters related to the sensory and anticipatory inputs, as well as network connectivity and architecture (Fig. S4-5). Increasing stimulus intensity led to a faster encoding of the stimulus in both conditions (Fig. S4a). Stimuli were always decoded faster when preceded by the cue (Fig. S4a; two–way ANOVA with factors 'stimulus slope', $p < 10^{-18}$, $F(4) = 30.4$, and 'condition' (expected vs. unexpected), $p < 10^{-15}$, $F(1) = 79.8$). The amount of anticipatory activity induced by the cue depended on stimulus intensity only weakly (p(interaction)= 0.05, $F(4) = 2.4$), and was present even in the case of step–like stimulus (Fig. S4b). Moreover, we found that anticipatory activity was maintained when we increased the number of presented stimuli, while keeping a fixed 50% probability that each cluster be selective to a given stimulus (Fig. S4c; two-way ANOVA with factors 'number of stimuli' ($p = 0.57$, $F(3) = 0.67$) and 'condition' ($p < 10^{-7}$, $F(1) = 35.3$; p(interaction)= 0.66, $F(3) = 0.54$). Anticipatory activity was present also when the stimulus selectivity targeted both $E$ and $I$ neurons, rather than $E$ neurons only (Fig. S4d; 50% of both $E$ and $I$ neurons were targeted by the cue).

We then tested robustness of anticipatory activity to variations in network size and architecture. We estimated the decoding accuracy in ensembles of constant size (20 neurons) sampled from networks of increasing size (Fig. S5a). When scaling the network size we kept fixed both the clusters' size and the probability that each cluster was selective to a given stimulus (50%). Network synaptic weights scaled as reported in Table 1 and 2 of Methods. Cue-induced anticipation was even more pronounced in larger networks (Fig. S5a, two–way ANOVA with factors 'network size' ($p < 10^{-20}$, $F(3) = 49.5$), 'condition' ($p < 10^{-16}$, $F(1) = 90$; p(interaction)= $10^{-10}$, $F(3) = 20$).

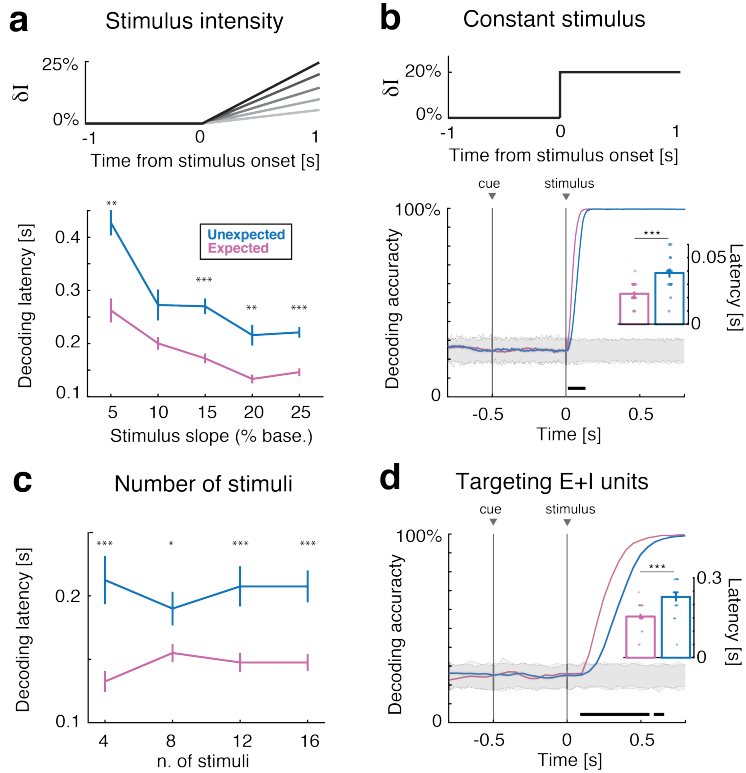

**Fig. S4**. **Robustness of anticipation effects to variations in stimulus model**. **a**: Stimulus intensity (slope). Latency of significant decoding increased with stimulus intensity in both conditions (top, darker shades represent stronger stimuli; bottom, decoding latency, mean±s.e.m.), and it is faster in expected (pink) than unexpected trials (blue). **b**: Step-like stimuli. Anticipatory activity is present in the case of step-like stimuli. *Bottom panel:* time course of decoding accuracy (notations as in Fig. 1c of the main text). **c**: Number of stimuli. Anticipatory activity did not depend on the number of stimuli presented to the network. **d**: Stimuli targeting inhibitory neurons. Anticipatory activity was present when stimuli targeted both $E$ and $I$ neurons. Notations as in bottom row of panel $a$. Panels **a-d**: $^{*} = p < 0.05$, $^{**} = p < 0.01$, $^{***} = p < 0.001$, post-hoc multiple–comparison $t$–test with Bonferroni correction. Horizontal black bar, $p < 0.05$, t–test with multiple–bin Bonferroni correction. *Insets*: $^{***} = p < 0.001$, signed–rank test.

In our main model the neural clusters were segregated in disjoint groups. In Fig. S5b, we investigated an alternative scenario where neurons may belong to multiple clusters, resulting in an architecture with overlapping cluster membership [3]. We found that anticipatory activity was present also in networks with overlapping clusters (Fig. S5b; see Sec. 2.4 of Supplementary Methods below for details of this model).

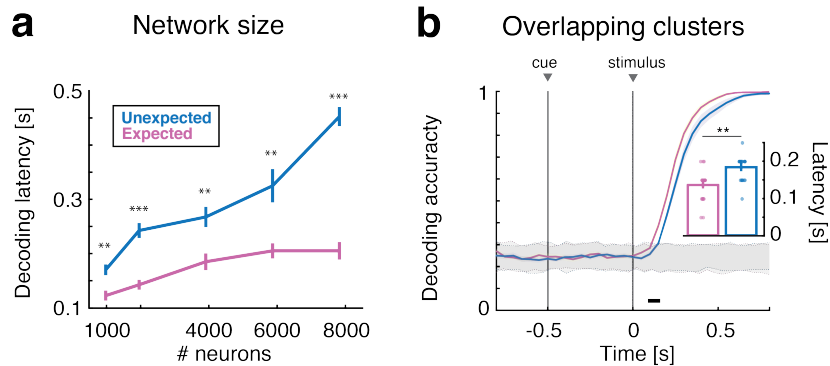

**Fig. S5**. **Robustness of anticipatory activity to variations in network architecture**. **a:** Variations in network size. Anticipatory activity was present for a large range of network sizes ($J_{+} = 5, 10, 20, 30, 40$ for $N = 1, 2, 4, 6, 8 \times 10^3$ neurons, respectively). **b:** Network with overlapping clusters. Anticipatory activity was present also when neurons were shared among multiple clusters (see the text for detail, notations as in Fig. 1c of the main text). Panels **a-b**: $^{**} = p < 0.01$, $^{***} = p < 0.001$, post-hoc multiple–comparison $t$–test with Bonferroni correction. Horizontal black bar, $p < 0.05$, $t$–test with multiple–bin Bonferroni correction. *Inset:* $^{**} = p < 0.01$, signed–rank test.

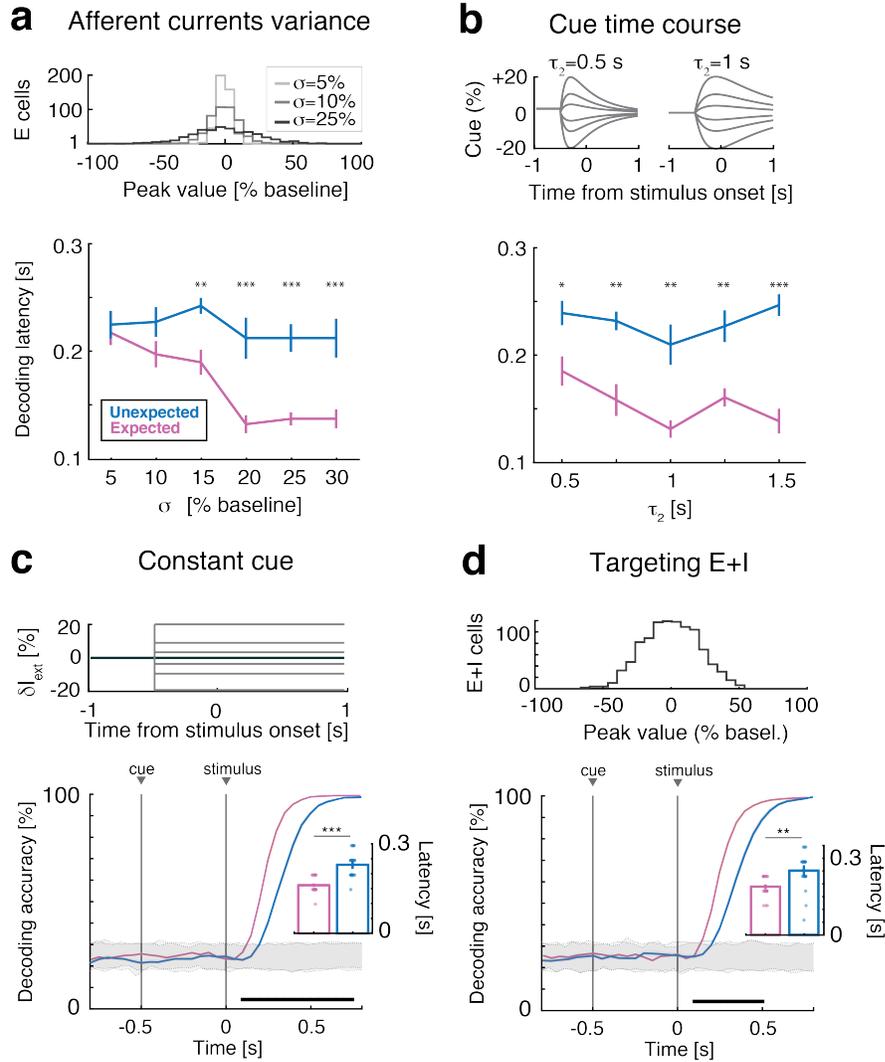

**Fig. S6**. **Robustness of anticipatory activity to variations in cue model**. **a**: Robustness to variations in spatial variance. Increasing the cue-induced spatial variance in the afferent currents $\sigma$ (top: distribution of peak afferent currents across neurons) leads to more pronounced anticipatory activity (bottom, latency in unexpected (red) and expected (pink) trials). **b**: Robustness to variations in kinetics (time course of cue stimulation). Anticipatory activity is present for a large range of cue time courses (top, double exponential cue profile with rise and decay times $[\tau_1, \tau_2] = g \times [0.1, 0.5]$ s, for $g$ in the range from 1 to 3; bottom, decoding latency during unexpected, red, and unexpected, pink, trials). **c**: Robustness to step–like kinetics. This example shows the case of step–like time course for the cue with spatial variance $\sigma = 20\%$. **d**: Robustness to inclusion of inhibitory neurons as targets. Anticipatory activity was also present when the cue targets 50% of E and I neurons ($\sigma = 20\%$ in baseline units). Panels **a-d**: $^{*} = p < 0.05$, $^{**} = p < 0.01$, $^{***} = p < 0.001$, post-hoc multiple–comparison t–test with Bonferroni correction. Horizontal black bar, $p < 0.05$, t–test with multiple–bin Bonferroni correction; Insets: $^{**} = p < 0.01$, $^{***} = p < 0.001$, signed–rank test. Panels **c-d**: notations as in Fig. 1c of the main text.

The results so far show that anticipatory activity was robust to changes in network architecture. Next, we show that anticipatory activity was robust to variations in cue parameters – specifically, (i) variations in the spatial variance $\sigma^2$ of the cue-induced afferent currents, (ii) variations in the kinetics of cue stimulation, and (iii) variations in the type of neurons targeted by the cue.

(i) We found that coding anticipation was present for all values of $\sigma$ above $10\%$ (Fig. S6a, two–way ANOVA with factors 'cue variance' ($p = 0.09$, $F(7) = 1.8$) and 'condition' ($p < 10^{-7}$, $F(1) = 31$); p(interaction)$= 0.07$, $F(7) = 1.9$).

(ii) Anticipatory activity was also robust to variations in the time course of the cue-evoked currents (Fig. S6b, two–way ANOVA with factors 'time course' ($p = 0.03$, $F(4) = 2.8$) and 'condition' ($p < 10^{-15}$, $F(1) = 72.7$; p(interaction)$= 0.12$, $F(4) = 1.8$). We also considered a model with constant cue-evoked currents (step-like model), to further investigate a potential role (if any) of the cue's variable time course on anticipatory activity (Fig. S6c). Again, we found anticipatory coding also in this case (Fig. S6c, bottom panel).

(iii) Finally, we tested whether or not anticipation was present if $I$ neurons, in addition to $E$ neurons, were targeted by the cue (while maintaining stimulus selectivity for $E$ neurons only). We assumed that 50% of both $E$ and $I$ neurons were targeted by the cue, and found robust anticipatory activity also in that case (Fig. S6d).

We did not find any anticipatory coding in the *homogeneous* network model as a result of the same manipulations of Fig. S4-S5 (not shown).

## 1.4 Distracting cue model

To show that the anticipatory effect of our model cue is specific, we give here an example of a manipulation leading to the opposite effect. Specifically, if the cue is modeled as recruiting recurrent inhibition rather than excitation (we refer to such a cue as a 'distractor'), stimulus decoding is slowed down rather than accelerated, as shown in Fig. S7a-b. The coding delay following such 'distracting cue' is due to a phenomenon of delayed dynamics which mirrors the anticipatory dynamics of the main model, as shown next.

First, from the analysis of model simulations we found that the activation latency of stimulus-selective clusters after stimulus presentation was longer during distracted trials (Fig. S7c), which resulted in the delayed onset of the coding states (see the main text for the definition of coding state in section *Anticipatory cue induces faster onset of taste-coding states*). This, in turn, was the result of higher energy barriers between coding and non-coding states, as confirmed in a mean field analysis of a simplified 2-cluster network (Fig. S7d). Except for the different cue targets, the 2-cluster network is the same as used to study anticipatory coding (main text, *Anticipatory cue speeds up the networks dynamics*). The distracting cue increased the energy barriers for a range of stimulus intensities (Fig. S7d, right panel).

## 1.5 Specificity of anticipatory mechanism (alternative models)

How specific is our model of anticipatory activity? To answer this question, we compared our model to several alternative models where the anticipatory cue modulates, respectively: the feedforward couplings $J_{ext}$; the recurrent couplings $J_{EE}$; and the background synaptic input.

All models had the same architecture of the main clustered network model and the same cue temporal profile, which was modeled as a double exponential with rise and decay times of 0.2s and 1s, respectively. In the alternative models, the cue targeted all clustered $E$ neurons, rather than just $50\%$ as in the main model of Fig. 1a, to avoid the introduction of spatial variance across neurons (this could confound the results with those of the main model). All models were scored on the ability to match the experimental data on the $\Delta$PSTH of cue responses and coding anticipation (see Fig. S8 and Table 1). The performance of the model of the main text is reported in Fig. S8a for comparison.

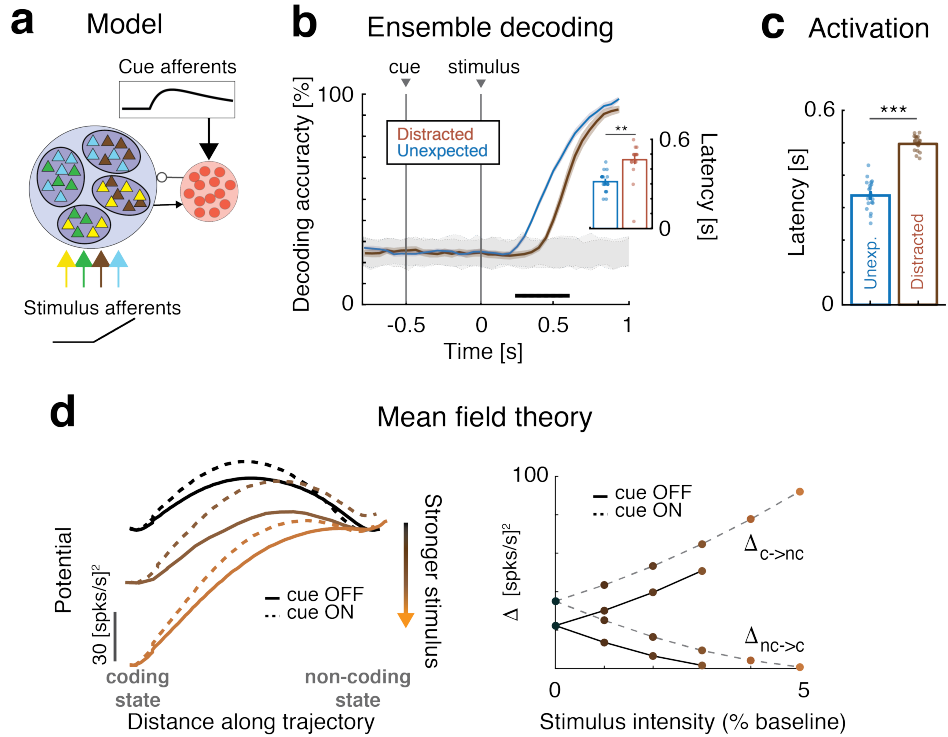

**Fig. S7. A "distracting" cue slows down stimulus coding (model).** **a**: Schematics of clustered network architecture and stimulation (notations as in Fig. 1a of the main text). Unlike the main model, here the cue targets the inhibitory neurons. **b**: Time course of cross–validated population decoding during distracted (brown) trials was slower than during unexpected (red) trials (notations as in Fig. 1c). *Inset:* aggregate analysis across 20 simulated sessions (mean±s.e.m.). **c**: Activation latency of stimulus-selective clusters after stimulus presentation was delayed during distracted trials (latency, mean±s.e.m.). **d**: Mean field theory of simplified 2-cluster network (notations as in Fig. 3c, lighter brown denotes stronger stimuli). Left panel: the transition probability from the non-coding (right well) to the coding state (left well) increased with larger stimuli. In distracted trials (dashed curves) the barrier height $\Delta$ from the non-coding to the coding state is larger than in unexpected trials (full curves), leading to slower coding in the distracted condition. Right panel: effective energy barriers as a function of stimulus intensity, with (full lines) and without the cue (dashed). Panels **b** and **c**: $^{**} = p < 0.01$, $^{***} = p < 0.001$, $t$-test.

*Feedforward coupling model* (Fig. S8b). In this model, The cue was a time dependent modulation of the external synaptic coupling $J_{ext} = J_{E0}$, identical for all clustered excitatory neurons. Coding anticipation was absent for either moderate or strong positive modulations varying from 10% to 20% above baseline. Cue responses were heterogeneous but to a different degree than the experimental data. For negative cue modulations, only inhibited cue responses were observed and no coding anticipation was present (not shown).

*Recurrent coupling model* (Fig. S8c). In this model, the cue was a time-dependent modulation of the recurrent coupling strength $J_{EE}$ between $E$ neurons. For negative modulation of the recurrent couplings (the more likely to produce a faster dynamics), coding anticipation was present, however, peak cue responses were

– 7 –

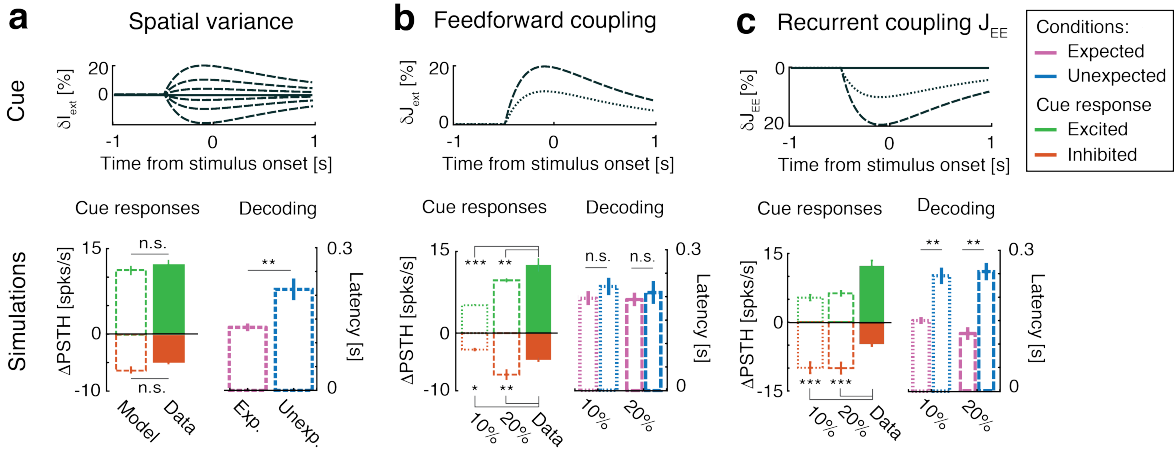

**Fig. S8. Alternative models. a**: Performance results of the main model ('spatial variance model' of the main text, see Fig. 1a-c), reported here for comparison. *Top row:* Temporal profile modeling the presentation of the cue (targeting excitatory neurons). *Bottom row, left:* ΔPSTH, defined as the peak cue response minus baseline rate, in data and model (excitatory responses in green, inhibited responses in red). *Bottom row, right:* Decoding performance in expected (pink) and unexpected (blue) conditions averaged across 20 simulated sessions. **b**: *Feedforward coupling model.* In this model, the cue elicited a modulation of the feedforward couplings $J_{ext}$ (see text). Same notations as in *a*. **c**: *Recurrent coupling model.* In this model, the cue elicited a modulation of *all* excitatory recurrent couplings $J_{EE}$ (see text). Same notations as in *a*. All panels: Network parameters as in Table 1. $t$-test, $^{*} = p < 0.05$, $^{**} = p < 0.01$, $^{***} = p < 0.001$.

strongly inhibited over a wide range of parameters, and thus incompatible with the empirical data. For positive modulation (not shown), coding anticipation was absent and cue responses were mostly excited.

*Background synaptic input model* (Fig. S9). In this model, the cue induced an increase in the background synaptic input, which in turn drove a simultaneous increase in background noise and shunting inhibition as in ref. [4]. The baseline noise level was modeled after an Ornstein-Uhlenbeck process with zero mean and variance $\sigma_{ext} = 0.5 r_{ext}$, where $r_{ext}$ was the mean afferent current. The cue increased the background noise by a factor $X$: $\sigma_{ext}^2 \to X \sigma_{ext}^2$ while shunting the membrane time constant by a factor $1/X$: $\tau_m \to \tau_m/X$ (we refer to [4] for details). Both modulations followed the same double exponential time course of the spatial variance model of the main text. With even a moderate factor of $X = 1.5$, the cue induced mostly inhibited cue responses and no anticipation (Fig. S9a), due to the strong shunting effect on the network dynamics, leading to a strong reduction of excitability in the clusters. At $X = 2$, excitatory neurons were transiently silenced (not shown). In an effort to obtain a fair comparison with the spatial variance model, we reduced the shunting effect (using peak value $\tau_m/X^\epsilon$, with $\epsilon = 1/2, 1/4$) while keeping the same amount of background noise (Fig. S9b). In this case, the cue induced heterogeneous cue responses more similar to the data, but still no anticipation of coding. We explored the parameter space by independently scaling background noise, shunting, and mean afferent current $r_{ext}$ (transiently increased with the same time course as the other quantities – to counteract the shunting effect), but found that coding anticipation was never present (Fig. S9c, bottom row). We concluded that a model cue inducing an increase in background synaptic activity did not lead to anticipatory activity.

In summary, while our spatial variance model produces robust coding anticipation and quantitatively matches cue responses in a wide range of parameters and architectures (Fig. S1, S4, S5, S6), all the potentially alternative models above failed to capture the empirical results and reproduce key features of the data (see Table 1 for a summary).

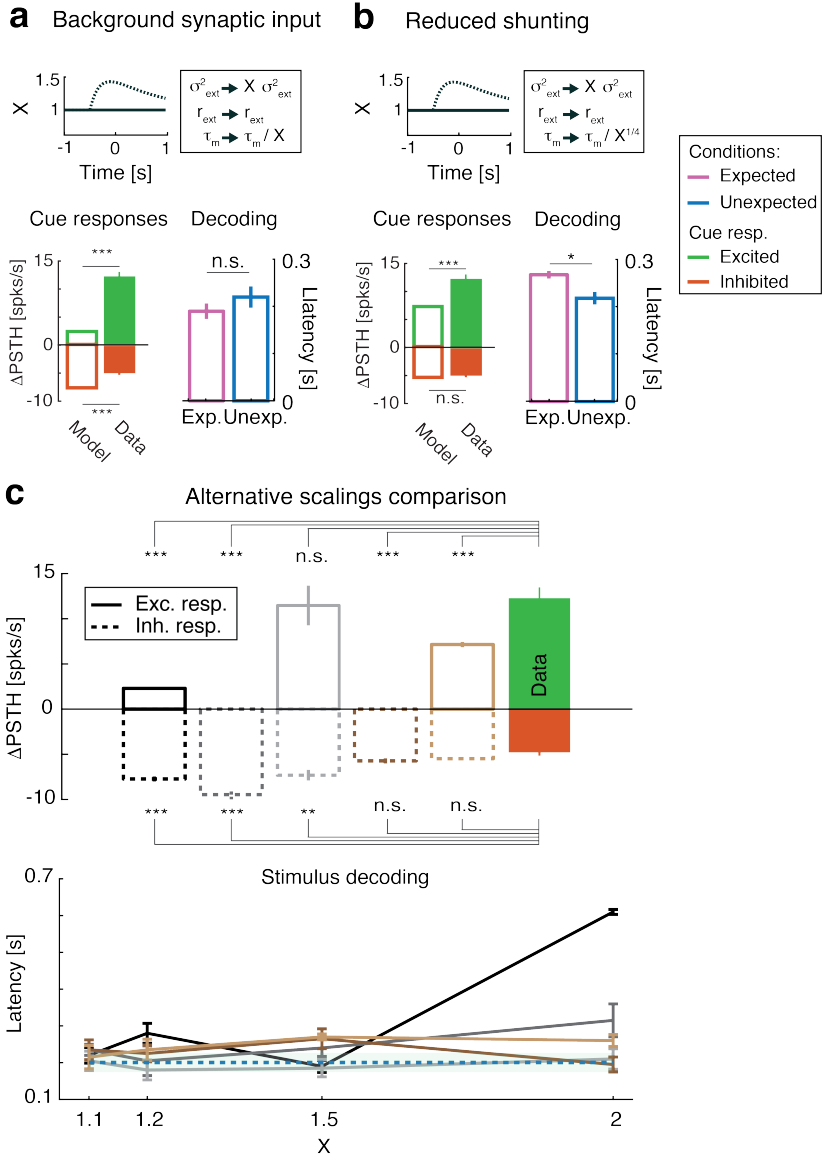

**Fig. S9. Background synaptic input model.** **a**: Main findings for the alternative model wherein the cue elicits a change in background synaptic input to $E$ neurons (see the text for details). Same notations as in Fig. 8a, with $X$: amount of cue-induced modulation ('scaling parameter'); $r_{ext}$: mean afferent current to the network; $\sigma_{ext}$: background noise input to the network; $\tau_m$: $E$ neurons membrane time constant. **b**: Model with reduced shunting inhibition. Same as $a$, with the membrane time constant scaled by a lower power of $X$. **c**: Effect of different scaling regimes on the performance of the background synaptic input model of panel $a$. *Top row:* $\Delta$PSTH. *Key:* Black, $\sigma^2_{ext} \to X\sigma^2_{ext}$, $r_{ext} \to r_{ext}$, $\tau_m \to \tau_m/X$ (same as panel **a**); Dark grey, $\sigma^2_{ext} \to X\sigma^2_{ext}$, $r_{ext} \to X^{1/2}r_{ext}$, $\tau_m \to \tau_m/X$; Light grey, $\sigma^2_{ext} \to X\sigma^2_{ext}$, $r_{ext} \to Xr_{ext}$, $\tau_m \to \tau_m/X$; Dark brown: $\sigma^2_{ext} \to X\sigma^2_{ext}$, $r_{ext} \to r_{ext}$, $\tau_m \to \tau_m/X^{1/2}$; Light brown: $\sigma^2_{ext} \to X\sigma^2_{ext}$, $r_{ext} \to r_{ext}$, $\tau_m \to \tau_m/X^{1/4}$. Rightmost bar: $\Delta$PSTH of data. *Bottom row:* decoding latency as a function of the scaling parameter $X$ for the 5 scaling regimes shown in top row (same color code). Dashed line: coding latency in the absence of the cue. All panels: $t$-test, $^* = p < 0.05$, $^{**} = p < 0.01$, $^{***} = p < 0.001$.

| | **Model** | | | |
|---|---|---|---|---|
| *Modulation* | Spatial Var. | Fwd. $J_{ext}$ | Rec. $J_{EE}$ | Backgr. Syn. Input |
| *Anticipation* | Yes (robust) | No | Yes | No |
| *Cue responses* | Yes (robust) | No | No | Yes (fine tuned) |
| *Figure* | S1, S4, S5, S6 | S8b | S8c | S9 |

**Table. S1**. Comparison of the performance of alternative models of the anticipatory cue (see Figs. S8, S9). First row ('Modulation') reports type of modulation due to the cue (feature defining the model; see the text). Second row ('Anticipation') reports the presence (Yes) or absence (No) of coding anticipation. Third row ('Cue responses') reports whether the $\Delta$PSTH is compatible (Yes) or not (No) with the empirical data. Last row: figure where the corresponding result is shown.

## 1.6 Firing rate coding of expectation

Stimulus coding could accelerate if the cue increased the firing rate of the stimulus-selective neurons compared to the non-selective neurons. Here we show that coding anticipation in the clustered network is *not* driven by such cue-induced changes in firing rate selectivity. To prove this point, we estimated the firing rate difference $\Delta r$ between stimulus-selective and nonselective neurons in the expected and unexpected conditions in the clustered network, and found no difference in $\Delta r$ (Fig. 10a).

Similarly, the finding that no coding acceleration is found in the homogeneous network (Fig. 1f of main text), suggests either no difference in $\Delta r$, or a difference in the opposite direction (i.e., larger $\Delta r$ in non-expected trials). An analysis of $\Delta r$ in the homogeneous network (Fig. 10b) confirmed that $\Delta r$ was larger in the unexpected condition, in agreement with the reversed trend in coding speed found in Fig. 1f.

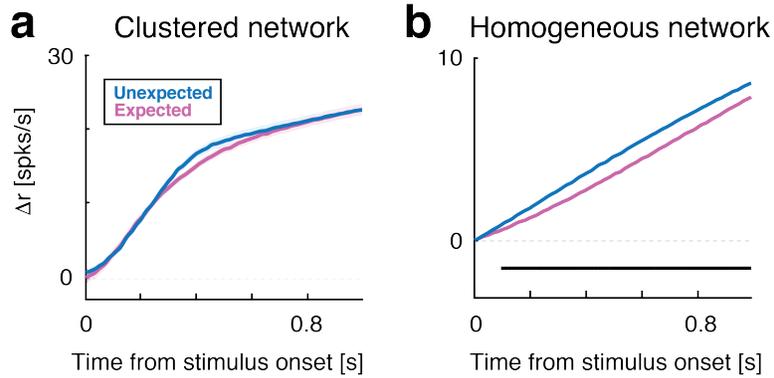

**Fig. S10**. Time course of the firing rate difference $\Delta r$ between stimulus-selective and nonselective clusters in the expected (pink) and unexpected (blue) conditions. **a:** Clustered network of Fig. 1a of main text. $\Delta r$ is not significantly different between expected and unexpected trials. **b:** Homogeneous network (Fig. 1d of main text). In this network, $\Delta r$ is significantly larger in the unexpected condition compared to the expected one (black horizontal bar: $p< 0.05$, $t$-test with multiple bin Bonferroni correction). Both panels: main curves represent means of $\Delta r$ over 20 sessions; shaded area represents s.e.m..

## Supplementary Methods

### 2.1 Behavioral training and electrophysiology

Adult female Long–Evans rats were used in the experiment. Movable bundles of 16 microwires attached to a "mini–microdrive" were implanted bilaterally in the gustatory cortex and intraoral cannulae (IOC) were inserted bilaterally and cemented. All experimental procedures were approved by the Institutional Animal Care and Use Committee of Stony Brook University and complied with university, state, and federal regulations on the care and use of laboratory animals (for more details, see [2]). After postsurgical recovery, rats were trained to self–administer fluids through IOCs by pressing a lever under head-restraint within 3s presentation of an auditory cue ("expected trials"; a 75 dB pure tone at a frequency of 5 kHz). The inter-trial interval was progressively increased to $40 \pm 3$s. Early presses were discouraged by the addition of a 2s delay to the inter-trial interval. During training and electrophysiology sessions, additional tastants were automatically delivered through the IOC at random times near the middle of the inter–trial interval and in the absence of the anticipatory cue ("unexpected trials"). The following tastants were delivered: 100 mM NaCl, 100 mM sucrose, 100 mM citric acid, and 1 mM quinine HCl. Water (50 $\mu$l) was delivered to rinse the mouth clean through a second IOC, 5s after the delivery of each tastant.

Multiple single–unit action potentials were amplified, bandpass filtered, and digitally recorded. Single units were isolated using a template algorithm, clustering techniques, and examination of inter–spike interval plots (Offline Sorter, Plexon). Starting from a pool of 299 single neurons in 37 sessions, neurons with peak firing rate lower than 1 spks/s (defined as silent) were excluded from further analysis, as well as neurons with a large peak around the 6–10 Hz in the spike power spectrum, which were classified as somatosensory [5–7]. Only ensembles with 5 or more simultaneously recorded neurons were included in the rest of the analyses. Ongoing and evoked activity were defined as occurring in the 5 seconds-long interval before or after taste delivery, respectively.

### 2.2 Cue responsiveness

A neuron was deemed responsive to the cue if a sudden change in its firing rate was observed during the post-cue interval. To detect changes in firing rates we used the 'change-point' (CP) procedure described in ref. [6] (an adaptation of a method presented in ref. [8]). Briefly, we built the cumulative distribution function of the spike count (CumSC) across all trials in each session, in the interval starting 0.5s before and ending 1 s after the cue delivery. We then ran the CP detection algorithm on CumSC record with a given tolerance level corresponding to a desired confidence level of $p = 0.05$ for a two-sample problem in spike counts. This algorithm looks for putative CPs as those points $x_C$ such that CumSC$(x_C)$ is located at the maximal distance from the straight line joining the first and any subsequent point $x_n$ along the record. If the spike counts before $x_C$ and between $x_C$ and $x_n$ are significantly different (according to the chosen tolerance level), $x_C$ is accepted as a legitimate change point. Normally, one would at this point repeat the procedure considering $x_C$ as the new start of the record, to look for additional CPs; however, we stopped the procedure after finding the first CP (if any), as this was sufficient to claim responsiveness. If, with this method, a CP was detected *before* cue presentation, the whole algorithm was repeated with lower tolerance (lower $p$-value) to enforce a more stringent criterion. If a legitimate CP was found anywhere within 1s after cue presentation, the neuron was deemed responsive. If no CP was found, the neuron was deemed not responsive. For neurons with excited (inhibited) cue responses, the positive (negative) peak response was estimated as the difference between the post-cue activity and the mean baseline activity in the 0.5s preceding cue presentation. This measure is reported as the ΔPSTH in e.g. Sec. 1.1 and Fig. S1.

## 2.3 Infinitesimal moments for mean field theory

Here we report the formulae for the infinitesimal mean and variance of the input current needed for mean field theory (see Methods, section *Mean field theory*).

We need the infinitesimal moments for $Q+2$ neural populations: the first $Q$ populations representing the $Q$ excitatory clusters; the $(Q+1)$–th population representing the "background" unstructured excitatory population; and the $(Q+2)$–th population representing the inhibitory population. The theory can be derived from a population density approach where the input to each neuron in population $\alpha$ is completely characterized by the infinitesimal mean $\mu_\alpha$ and variance $\sigma_\alpha^2$ of the post-synaptic current (see Table 1 of Methods for parameter values). The mean and variance of the infinitesimal moments of the input current to the $\alpha^{th}$ excitatory population is

$$\mu_\alpha = \tau_{m,E}\sqrt{N}\Big(\frac{n_E f}{Q}\Big(p_{EE}J_+ j_{EE} r_\alpha + \sum_{\beta=1}^{Q-1} p_{EE}J_- j_{EE} r_\beta\Big) + n_E(1-f)p_{EE}J_- j_{EE} r_E^{(bg)}$$
$$-n_I p_{EI} j_{EI} r_I + n_E p_{E0} j_{E0} r_{\text{ext}}\Big),$$

$$\sigma_\alpha^2 = \tau_{m,E}\Big(\frac{n_E f}{Q}\Big(p_{EE}(J_+ j_{EE})^2(1+\delta^2) r_\alpha + \sum_{\beta=1}^{Q-1} p_{EE}(J_- j_{EE})^2(1+\delta^2) r_\beta\Big)$$
$$+n_E(1-f)p_{EE}(J_- j_{EE})^2(1+\delta^2) r_E^{(bg)} + n_I p_{EI} j_{EI}^2(1+\delta^2) r_I\Big),$$

where $r_\alpha$, $r_\beta$, with $\alpha, \beta = 1, ..., Q$, are the firing rates of the excitatory clusters; $r_E^{(bg)}$ is the firing rate of the background $E$ population; $r_I$ is the firing rate of the $I$ population; $n_E = 4/5$ and $n_I = 1/5$ are the fractions of excitatory and inhibitory neurons in the network, respectively, and $\delta = 1\%$ is the SD of the distribution of synaptic weights. The infinitesimal moments of the afferent current to a neuron belonging to the background $E$ population read

$$\mu_E^{(bg)} = \tau_{m,E}\sqrt{N}\Big(\frac{n_E f}{Q}\sum_{\beta=1}^{Q} p_{EE}J_- j_{EE} r_\beta + n_E(1-f)p_{EE} j_{EE} r_E^{(bg)}$$
$$-n_I p_{EI} j_{EI} r_I + n_E p_{E0} j_{E0} r_{\text{ext}}\Big),$$

$$(\sigma_E^{(bg)})^2 = \tau_{m,E}\Big(\frac{n_E f}{Q}\sum_{\beta=1}^{Q} p_{EE}(J_- j_{EE})^2(1+\delta^2) r_\beta + n_E(1-f)p_{EE} j_{EE}^2(1+\delta^2) r_E^{(bg)}$$
$$+n_I p_{EI} j_{EI}^2(1+\delta^2) r_I\Big),$$

and, similarly, for the current to a neuron of the inhibitory population we have

$$\mu_I = \tau_{m,I}\sqrt{N}\Big(\frac{n_E f}{Q}\sum_{\beta=1}^{Q} p_{IE}j_{IE}r_\beta + n_E(1-f)p_{IE}j_{IE}r_E^{(bg)}$$
$$-n_I p_{II}j_{II}r_I + n_E p_{I0}j_{I0}r_{\text{ext}}\Big),$$
$$\sigma_I^2 = \tau_{m,I}\Big(\frac{n_E f}{Q}\sum_{\beta=1}^{Q} p_{IE}j_{IE}^2(1+\delta^2)r_\beta + n_E(1-f)p_{IE}j_{IE}^2(1+\delta^2)r_E^{(bg)}$$
$$+n_I p_{II}j_{II}^2(1+\delta^2)r_I\Big). \tag{2.1}$$

These expressions enter the mean field equations as described in Methods (section *Mean field theory*).

## 2.4 Network with overlapping clusters

In the spiking network with overlapping clusters of Fig. S5b, each neuron (network size: $N = 2000$) had a probability $f = 0.06$ of belonging to one of the $Q = 14$ clusters [3]. Hence, the fraction of neurons in a population shared by a set of $k$ specific clusters (a population of 'rank' $k$) was $f^k(1-f)^{Q-k}$ (there are $\binom{Q}{k}$ such populations); and the fraction of neurons in the unstructured background population (rank 0) was $(1-f)^Q$. Synaptic weights $J_{ij}$ between pre- and post-synaptic excitatory neurons $j$ and $i$ respectively, were given by the stochastic variable

$$J_{ij} = p_{EE}\left(\epsilon_{ij}\xi_+ + (1-\epsilon_{ij})\xi_-\right),$$

where $p_{EE} = 0.2$ is the $E \to E$ connection probability, $\xi_\pm$ are the two possible values of the synaptic weights (see below), and $\epsilon_{ij}$ and $1-\epsilon_{ij}$ are the potentiation and depression probabilities, $P(J_{ij} = \xi_+)$ and $P(J_{ij} = \xi_-)$, respectively. Weights were potentiated or depressed according to the rule [3, 9]

$$\epsilon_{ij} = \frac{P_{ij}}{P_{ij} + \rho f D_{ij}}, \tag{2.2}$$

where $P_{ij} = \sum_{k=1}^{Q}\eta_i^k\eta_j^k$ is the number of clusters in common between neurons $i$ and $j$ ($\eta_i^k = 1$ if cluster $k$ contains neuron $i$ and $\eta_i^k = 0$ otherwise), while $D_{ij} = \sum_{k=1}^{Q}\eta_i^k(1-\eta_j^k)$, and $\rho = 2.75$. To motivate Eq. 2.2, we note that this is the asymptotic value of the potentiation probability obtained from the stochastic Hebbian learning rule of ref. [9] in the limit of slow learning. Finally, $\xi_+$ and $\xi_-$ were random variables sampled from normal distributions with means $J_+$ and $J_-$ and variances $\delta^2 J_+^2$ and $\delta^2 J_-^2$, respectively ($\delta = 0.01$). The $E \to I$, $I \to E$, and $I \to I$ connection probability was the same as for the homogeneous and clustered networks described (Table 1 in the main text); the stimuli and the anticipatory cue targeted the excitatory neurons as in the main model with segregated clusters. The remaining parameters of the network are reported in Table S2.

| Symbol | Description | Value |
|---|---|---|
| $j_{EE}$ | Mean E→E weights $\times \sqrt{N}$. | 1.4 mV |
| $j_{EI}$ | Mean I→E weight $\times \sqrt{N}$. | 5.0 mV |
| $j_{IE}$ | Mean E→I weight $\times \sqrt{N}$. | 2.5 mV |
| $j_{II}$ | Mean I→I weight $\times \sqrt{N}$. | 6.1 mV |
| $j_{E0}$ | Mean afferent synaptic weights to E neurons $\times \sqrt{N}$. | 7.3 mV |
| $j_{I0}$ | Mean afferent synaptic weights to I neurons $\times \sqrt{N}$. | 6.5 mV |
| $J_+$ | Potentiated intra-cluster E→E weights factor. | 10.5 |
| $r_{ext}^E$ | Average afferent rate to E neurons (baseline). | 5 spks/s |
| $r_{ext}^I$ | Average afferent rate to I neurons (baseline). | 7 spks/s |
| $V_E^{thr}$ | E neuron threshold potential. | 3.6 mV |
| $V_I^{thr}$ | I neuron threshold potential. | 5.7 mV |
| $V^{reset}$ | E and I neurons reset potential. | 0 mV |
| $\tau_m$ | E and I membrane time constant. | 20 ms |
| $\tau_{ref}$ | Absolute refractory period. | 5 ms |
| $\tau_{syn}$ | E and I synaptic time constant. | 4 ms |

**Table. S2**. Parameters for the network with overlapping clusters with $N = 2000$ LIF neurons. See Sec. 2.4 for details.